\documentclass[showkeys,aip,graphicx,amsmath,superscriptaddress,amssymb]{revtex4-2}
\usepackage[a4paper,margin=1in]{geometry}

\usepackage{CJKutf8}
\usepackage{lineno,hyperref}
\usepackage[normalem]{ulem}
\usepackage{graphics}
\usepackage{amsmath,bm}
\usepackage{amssymb}
\usepackage{graphicx}
\usepackage{bm}
\usepackage{amsfonts}
\usepackage{algorithm}
\usepackage{color}
\usepackage{epstopdf}
\usepackage{booktabs}
\usepackage{array}
\usepackage{hyperref}
\usepackage{graphics}
\usepackage{epsfig}
\usepackage{algorithm}
\usepackage{algpseudocode}
\usepackage{bm}
\usepackage{epic,eepic}
\usepackage{epsfig}
\usepackage{pifont}
\usepackage{braket}
\usepackage{subfigure}

\usepackage{xspace}
\usepackage{graphicx,color}
\usepackage{amsmath,amssymb,amsfonts,mathrsfs}
\usepackage[normalem]{ulem}
\usepackage{booktabs}
\usepackage{float}
\usepackage[toc,page]{appendix}
\usepackage[capitalize]{cleveref}
\usepackage{mathtools}
\usepackage[normalem]{ulem}
\usepackage{gensymb}
\usepackage{lineno}
\usepackage{hyperref}

\begin{document}

\title{Uni-Flow: a unified autoregressive-diffusion model for complex multiscale flows}

\author{Xiao Xue}
\affiliation{ 
Centre for Computational Science, University College London, London, UK
}
\author{Tianyue Yang$^{\dagger}$}
\affiliation{ 
Centre for Computational Science, University College London, London, UK
}
\author{Mingyang Gao$^{\dagger}$}
\affiliation{ 
Centre for Computational Science, University College London, London, UK
}
\author{Leyu Pan}
\affiliation{ 
Department of Earth Science and Engineering, Imperial College London, London, UK
}
\author{Maida Wang}
\affiliation{ 
Centre for Computational Science, University College London, London, UK
}
\author{Kewei Zhu}
\affiliation{ 
Department of Chemical Engineering, University College London, London, UK
}
\author{Shuo Wang}
\affiliation{ 
Department of Physics, Eindhoven University of Technology, Eindhoven, Netherlands
}
\author{Jiuling Li}
\affiliation{ 
School of Civil \& Environmental Engineering, 
Queensland University of Technology, Brisbane, Australia 
}
\affiliation{ 
Australian Centre for Water and Environmental Biotechnology, The University of Queensland, Brisbane, Australia
}
% \author{Tianyi Li}
% \affiliation{ 
% Department of Physics, University of Rome Tor Vergata, Rome, Italy. 
% }
\author{Marco F.P. ten Eikelder}
\affiliation{Institute for Mechanics, Computational Mechanics Group, Technical University of Darmstadt, Germany}
\author{Peter V. Coveney}%
\email{p.v.coveney@ucl.ac.uk}
\affiliation{ 
Centre for Computational Science, University College London, London, UK
}
\affiliation{
Centre for Advanced Research Computing, University College London, London, UK
}
\thanks{$^{\dagger}$These authors contributed equally as second authors.}

\date{\today}

%\maketitle
\begin{abstract}
Spatiotemporal flows govern diverse phenomena across physics, biology, and engineering, yet modelling their multiscale dynamics remains a central challenge. Despite major advances in physics-informed machine learning, existing approaches struggle to simultaneously maintain long-term temporal evolution and resolve fine-scale structure across chaotic, turbulent, and physiological regimes. Here, we introduce Uni-Flow, a unified autoregressive–diffusion framework that explicitly separates temporal evolution from spatial refinement for modelling complex dynamical systems. The autoregressive component learns low-resolution latent dynamics that preserve large-scale structure and ensure stable long-horizon rollouts, while the diffusion component reconstructs high-resolution physical fields, recovering fine-scale features in a small number of denoising steps. We validate Uni-Flow across canonical benchmarks, including two-dimensional Kolmogorov flow, three-dimensional turbulent channel inflow generation with a quantum-informed autoregressive prior, and patient-specific simulations of aortic coarctation derived from high-fidelity lattice Boltzmann hemodynamic solvers. In the cardiovascular setting, 
Uni-Flow enables task-level faster than real-time inference of pulsatile hemodynamics, reconstructing high-resolution pressure fields over physiologically relevant time horizons in seconds rather than hours. 
% Uni-Flow enables task-level faster-than-real-time inference of pulsatile hemodynamics by operating on a reduced spatiotemporal representation aligned with task-level observables.  
By transforming high-fidelity hemodynamic simulation from an offline, HPC-bound process into a deployable surrogate, Uni-Flow establishes a pathway to faster-than-real-time modelling of complex multiscale flows, with broad implications for scientific machine learning in flow physics.

% \XX{
% 1. machine precision\\
% 2. Figure 1 update\\
% 3. Missing formal metrics ?\\
% 4. QIML looks sudden\\
% 5. Diffusion model description is too long, but key details missing\\
% 6. Computational cost comparison requires standardisation
% Figure 1 update
% NS2D numerical solver speed
% GPU specification
% 2. Add COA boundary condition. \& error table\\
% 3. Title: high resolution, check NMI\\
% 4. why our model? what's the difference between current and other timeseries results\\
% 5. mention mask, mention latent diffusion (DiT), mention our advantage\\
% }

\end{abstract}

\maketitle

%\maketitle %\maketitle must follow title, authors, abstract and \pacs
\section{INTRODUCTION}

Partial differential equations (PDEs) govern the spatiotemporal evolution of many dynamical systems central to science and engineering. From canonical problems in fluid dynamics to physiologically critical processes such as cardiovascular flows~\cite{mazzeo2008hemelb}, these systems often exhibit nonlinear, turbulent, and chaotic behaviour~\cite{coveney2025molecular}. Accurately capturing and computing such dynamics across disparate temporal and spatial scales is essential for advancing both fundamental understanding and practical applications~\cite{evans2022partial}. Traditionally, PDEs are solved numerically using finite difference~\cite{strikwerda2004finite}, finite volume~\cite{eymard2000finite}, and finite element~\cite{reddy1993introduction} discretisations, which have enabled decades of progress in computational physics and engineering. Despite these advances, resolving multiscale dynamics remains computationally prohibitive, increasingly relying on exascale computing resources~\cite{atchley2023frontier} and facing fundamental challenges arising from the simultaneous presence of long-term evolution and fine-scale structure~\cite{hughes1995multiscale}. In computational fluid dynamics (CFD), large eddy simulation (LES)~\cite{mason1994large,piomelli1999large,xue2022synthetic} mitigates some of this cost by filtering unresolved small-scale motions. However, LES remains prohibitively expensive for many academic and industrial applications due to the fine grid resolution required near solid boundaries~\cite{yang2021grid}, limiting its practicality for inverse design~\cite{shirvani2023machine} and time-sensitive simulations. These computational constraints are particularly restrictive in physiological flow modelling, where high-fidelity solvers remain largely confined to offline analysis.

Scientific machine learning (SciML) has emerged as a promising paradigm for accelerating the modelling of PDE-governed systems by learning their evolution directly from data~\cite{brunton2020machine}. Recent advances have demonstrated substantial speedups in simulating complex dynamical systems~\cite{li2020fourier,xue2025equivariant}, especially in regimes where traditional solvers are computationally demanding. Rather than discarding physical structure, physics-informed machine learning (PIML) incorporates conservation laws~\cite{karniadakis2021physics,cheng2025machine}, symmetries~\cite{satorras2021n}, and governing equations into the learning process, ensuring physical consistency while leveraging the expressive capacity of modern machine learning models. Within CFD, machine learning has been applied to enhance PDE modelling through subgrid-scale closures~\cite{hoekstra2026reduced,van2025energy} and turbulence modelling~\cite{beck2019deep,duraisamy2019turbulence,sarghini2003neural,gamahara2017searching}, where resolving all flow scales is impractical. Reinforcement learning has been used to adaptively tune closure terms~\cite{xie2019artificial,mohan2023embedding,fischer2025optimal}, while physics-informed neural networks embed PDE residuals directly into loss functions~\cite{karniadakis2021physics}. These approaches reflect a broader trend toward hybridising data-driven learning with numerical solvers, allowing machine learning models to respect governing physical constraints.

Beyond targeted model enhancement, recent work has focused on learning the spatiotemporal evolution of entire fields. Recurrent neural architectures~\cite{sherstinsky2020fundamentals}, such as long short-term memory (LSTM) networks~\cite{greff2016lstm}, capture temporal dependencies but often struggle with long-horizon stability. Neural operator frameworks, including the Fourier Neural Operator~\cite{li2020fourier} and DeepONet~\cite{lu2021learning}, provide a general approach for learning mappings between infinite-dimensional function spaces, enabling efficient solutions of parametric PDEs. Despite their promise, most autoregressive and operator-learning models suffer from long-term instability, where accumulated errors degrade physical realism or cause predictions to collapse towards trivial states. Efforts to stabilise chaotic dynamics through invariant-measure preservation and constrained learning~\cite{kochkov2021machine,ye2025recurrent,wangxue2025quantum} have largely been limited to simplified or low-resolution systems. Despite these advances, in general, such approaches have struggled to preserve the correct macroscopic dynamics, with machine-learned surrogate models often diverging from the true long-term behaviour~\cite{mccabetowards}.

A complementary line of work has explored diffusion models~\cite{song2020score,ho2020ddpm,mokady2022nulltextinversioneditingreal} for reconstructing high-resolution physical fields. Originally developed for image synthesis and restoration~\cite{esser2021tamingtransformer}, diffusion models have demonstrated strong performance in recovering fine-scale structure and multiscale features in complex flows~\cite{lam2023learning}. However, diffusion-based approaches lack explicit mechanisms for temporal evolution, making long-horizon prediction and causal consistency difficult to maintain~\cite{ruhling2023dyffusion}. Latent diffusion formulations further compress physical states into low-dimensional representations~\cite{rombach2022high,du2024conditional}, reducing interpretability and potentially obscuring underlying physics. In addition, iterative denoising procedures remain computationally demanding, limiting their suitability for real-time or long-duration simulations. These developments expose a fundamental tension in data-driven modelling of multiscale dynamical systems: models optimised for stable long-term temporal evolution typically operate at coarse resolution and struggle to recover fine-scale structure, while models capable of high-resolution spatial reconstruction often lack explicit temporal dynamics. This conflict is particularly pronounced in physiological flows, where both long-horizon consistency and fine-scale accuracy are essential.

\begin{figure}[htbp!]
\centering
\includegraphics[width=1\textwidth]{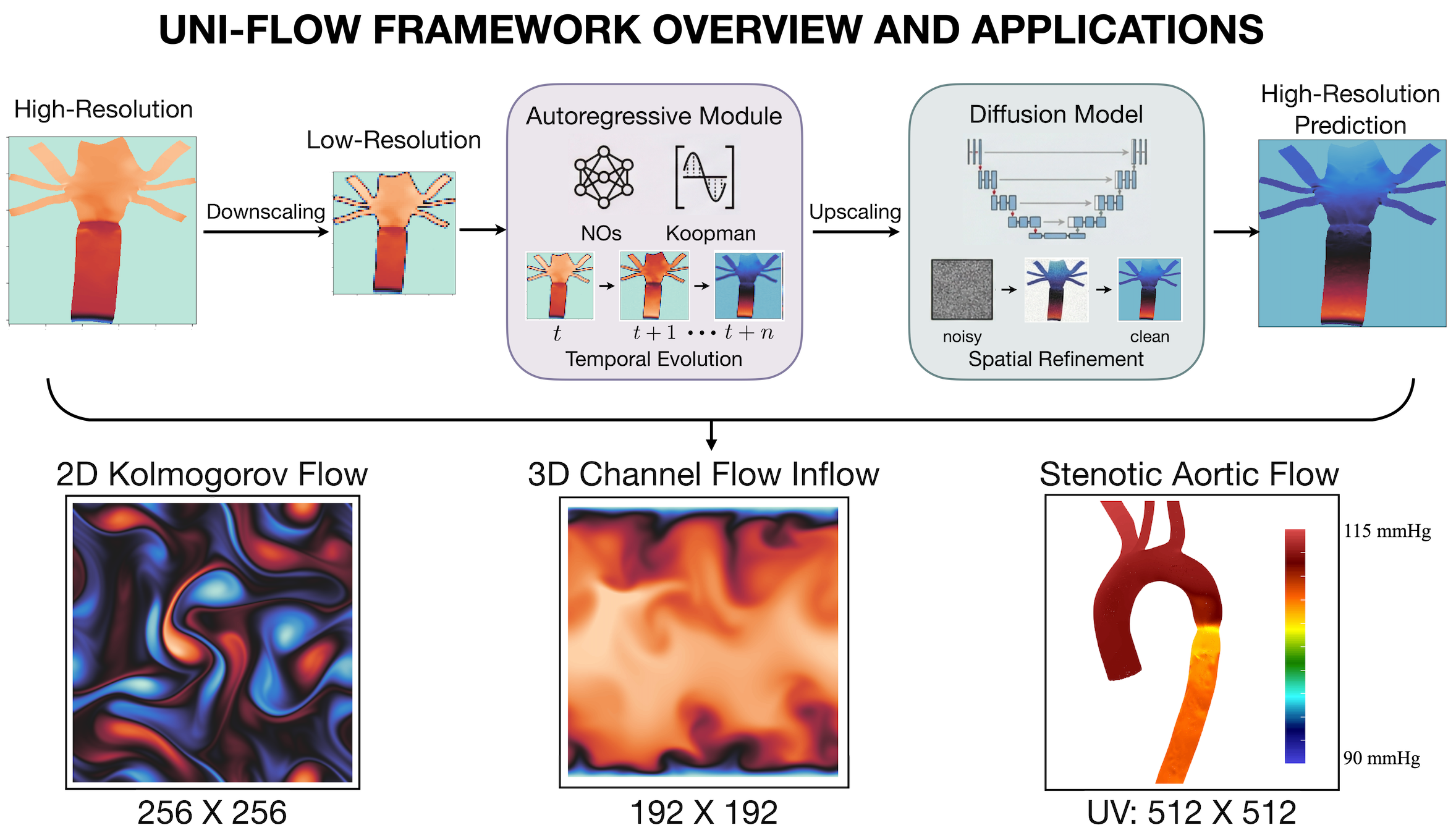}
\caption{\textbf{Uni-Flow framework for multiscale spatiotemporal modelling and representative applications.} Uni-Flow decouples temporal evolution and spatial refinement through a unified autoregressive–diffusion formulation. High-resolution physical fields are first downsampled to a low-resolution representation, 
% which is evolved forward in time by an autoregressive (AR) module operating in a reduced latent space. 
that retains the large-scale degrees of freedom governing long-horizon temporal evolution relevant to the target observables. 
The AR component may be instantiated using different operator-learning architectures, such as neural operators or Koopman-based models, to ensure stable long-horizon temporal dynamics. At selected timesteps, the low-resolution state is upscaled and provided as conditioning to a diffusion-based refinement model, which reconstructs high-resolution physical fields and recovers fine-scale spatial structure through a small number of denoising steps. The lower panels illustrate representative applications of Uni-Flow across increasing physical and modelling complexity: two-dimensional Kolmogorov flow ($256 \times 256$), three-dimensional turbulent channel inflow generation ($192 \times 192$), and patient-specific stenotic aortic flow, where surface pressure fields are parameterised on a two-dimensional UV domain ($512 \times 512$).}
\label{fig:sketch}
\end{figure}

Here, we introduce Uni-Flow, a unified autoregressive-diffusion framework designed around a deliberate separation of temporal evolution and spatial refinement (see Fig.~\ref{fig:sketch}). Uni-Flow explicitly decomposes these roles: a physics-informed autoregressive component evolves low-resolution states that preserve large-scale structure and long-term dynamics, while a diffusion-based refinement operator reconstructs high-resolution physical fields conditioned on the evolving latent state. By decoupling temporal stability from spatial fidelity, Uni-Flow avoids the long-term drift common to autoregressive models and the lack of causal consistency inherent to purely diffusion-based approaches. This formulation defines a general modelling principle for multiscale dynamical systems and is independent of specific neural architectures involved in machine learning approaches. We evaluate Uni-Flow across both canonical and physiologically relevant systems, progressing from controlled turbulence benchmarks to patient-specific cardiovascular simulations. These include two-dimensional Kolmogorov flow at Reynolds number $Re=|u|L_x/\nu=1000$ where $L_x$ is the characteristic length, $nu$ is the kinematic visocity, three-dimensional turbulent channel inflow generation, and human pulsatile aortic flow with coarctation, where a masked diffusion strategy is employed to recover high-resolution surface pressure fields. In the turbulent channel-flow case, the low-resolution autoregressive component is guided by a quantum-informed prior trained on a 20-qubit IQM device~\cite{wangxue2025quantum}, illustrating the framework’s ability to accommodate emerging learning paradigms without constraining its general formulation. Across all systems, Uni-Flow evolves the dynamics predominantly at low resolution and selectively applies diffusion-based refinement at chosen timesteps, enabling efficient long-horizon forecasting with adaptive transitions to high-resolution fields. 

For the hemodynamic demonstrations, we build directly on HemeLB~\cite{mazzeo2008hemelb,zacharoudiou2023development}, an exascale lattice Boltzmann solver designed for large-scale blood flow simulations, which provides the validated numerical foundation for our datasets and has been deployed on the Frontier supercomputer to generate high-fidelity pulsatile aortic flow data~\cite{atchley2023frontier}. In this setting, Uni-Flow enables faster than real-time inference of physiologically relevant hemodynamic evolution while reconstructing high-resolution pressure fields using an efficient denoising diffusion implicit model (DDIM) with only a small number of denoising steps. 

These results demonstrate that Uni-Flow consistently outperforms the aforementioned state-of-the-art neural operators and autoregressive baselines in both physical fidelity and computational efficiency. Importantly, Uni-Flow establishes a pathway towards real-time surrogate modelling of multiscale dynamical systems across physics, biology, and engineering, with particular relevance to time-sensitive applications in computational hemodynamics. In this context, such surrogates have the potential to reduce simulation turnaround times from days to seconds while preserving the correct pressure-drop characteristics associated with aortic stenosis.~\cite{norgaard2014diagnostic,tanade2022analysis}. Finally, Uni-Flow illustrates how generative flow modelling can be integrated with emerging quantum machine learning approaches within a unified computational framework.

\section{Applications }\label{sec:main}

Building on the hybrid autoregressive-diffusion framework, we evaluate Uni-Flow across three representative flow systems that progressively increase in physical and modelling complexity. We demonstrate the framework's ability to sustain long-horizon dynamics while selectively reconstructing fine-scale structure, a capability essential for practical scientific simulations. The two-dimensional Kolmogorov flow first serves as a canonical benchmark for statistically stationary chaotic flow. We then extend the framework to a inflow generator for three-dimensional wall-bounded turbulent channel flow, a canonical challenge in engineering~\cite{lund1998generation,wu2017inflow,xue2022synthetic}. Finally, the stenotic aortic flow highlights Uni-Flow’s potential for time-sensitive physiological modelling, where rapid and physically accurate predictions are essential. Across these regimes, Uni-Flow consistently outperforms autoregressive baselines in reconstruction accuracy, temporal stability, and physical realism, while enabling substantial reductions in computational turnaround time, reducing multi-hour, multi-GPU hemodynamic simulations to task-level observable inference on a single GPU, achieving faster-than-real-time generation of surface pressure fields~\cite{tanade2022analysis,coveney2025digital}. All experiments across the three application cases are conducted using single precision floating-point 32 bit (FP32) numbers.

\subsection{2D Kolmogorov flow}
% Problem introduction
The two-dimensional (2D) Kolmogorov flow provides a canonical testbed for statistically stationary turbulence, where sinusoidal forcing drives a self-organised array of counter-rotating vortices. Despite its simple formulation, the flow can exhibit complex interactions among coherent structures, energy cascades, and small-scale intermittency, making it an ideal benchmark for assessing turbulence models.
% Mathematical formulation
The two-dimensional Kolmogorov flow, governed by the 2D incompressible Navier-Stokes equation:
\begin{equation}
\frac{\partial \mathbf{u}}{\partial t} + (\mathbf{u} \cdot \nabla) \mathbf{u} = -\nabla p + \nu \Delta \mathbf{u} + \mathbf{F},
\end{equation}

\begin{equation}
\nabla \cdot \mathbf{u} = 0,
\end{equation}
where $\mathbf{u}$ is the velocity field and $p$ represents the pressure. $\nu$ is the kinematic viscosity, and the forcing term is given by $\mathbf{F} = 0.1[\sin(2\pi(x+y)) + \cos(2\pi(x+y))]$ with amplitude $0.1$. The domain is taken to be doubly periodic: $(x, y) \in [0, L_x] \times [0, L_y]$, with periodic boundary conditions in both $x$ and $y$ directions.

\begin{figure}[htbp!]
\centering
\includegraphics[width=1\textwidth]{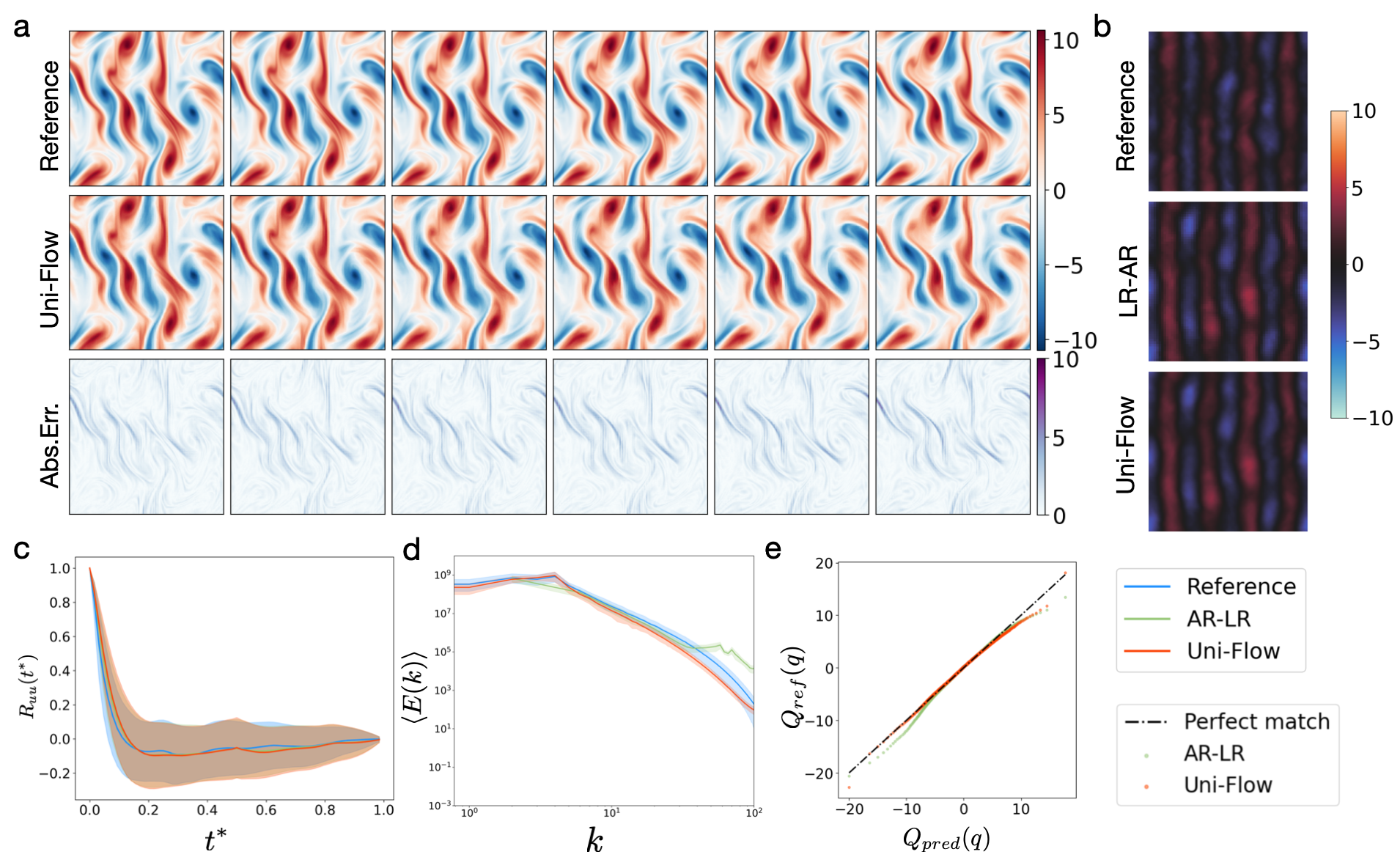}
%\label{fig:periodic-stg}
\caption{\textbf{Evaluation of the Uni-Flow model on the two-dimensional Kolmogorov flow.} Panel (a) shows sequential vorticity snapshots for the ground truth (top), low-resolution autoregressive (LR-AR) prediction (middle), and absolute error (bottom), where LR-AR captures the dominant shear-layer dynamics but under-resolves fine-scale vortices. Panel (b) presents the time-averaged vorticity field, showing that the LR-AR model reproduces the large-scale roll structures imposed by the Kolmogorov forcing. Panel (c) displays the temporal autocorrelation, indicating stable short-term dynamics with gradual decay at longer horizons. 
Panel (d) shows the time-averaged kinetic energy spectrum $\langle E(k) \rangle$, demonstrating consistent inertial scaling but slight underestimation at high wavenumbers. Panel (e) presents the Q-Q comparison between predicted and ground-truth vorticity distributions, confirming statistical alignment with minor deviation in the tails. }
\label{fig:KF_statistics}
\end{figure}
% Scenarios / cases
We generated a dataset consisting of 320 independent trajectories at a Reynolds number of $Re=1000$, each containing 320 temporal frames at a spatial resolution of $256\times256$. The dataset was divided into 80\% for training, 10\% for validation, and 10\% for testing. The detailed information regarding the dataset generation is demonstrated in Supplementary Information Section S1.1. The low-resolution training set is generated by uniformly subsampling, resulting in $64\times64$ grids that retain the dominant low wavenumber components responsible for large-scale flow structures. 
The high-resolution reconstruction is performed by 
%a diffusion model, guided through a DDIM process, reconstructing $256\times256$ flow fields from their $64\times64$ counterparts within 40 denoising steps. 
a diffusion-based refinement process that operates under the constraint imposed via low-resolution autoregressive dynamics, recovering fine-scale structure within 40 denoising steps. 
% Objective: State the learning task clearly: what is being learned.
The goal is to advance the vorticity field to the next timestep $t+1$, given its current state $\omega_t$.

% Results and interpretation
Figure~\ref{fig:KF_statistics} presents the performance of the low-resolution autoregressive model on the two-dimensional Kolmogorov flow compared with the ground-truth simulation.
Figure~\ref{fig:KF_statistics}\textbf{a} shows snapshots of the vorticity field over sequential time steps for the ground truth (top), Uni-Flow prediction (middle), and the corresponding absolute error (bottom) (See definition in Supplementary Information S6.5). The Uni-Flow model captures both large-scale shear patterns and fine scale vorticity structures and their temporal evolution. Figure~\ref{fig:KF_statistics}\textbf{b} presents the time-averaged vorticity field, indicating that the LR-AR baseline captures the overall mean structure, whereas Uni-Flow produces a smoother representation of the fine-scale variations. In all following cases, blue, green, and red denote the reference, LR-AR, and Uni-Flow. The represented color band denotes the standard deviation measured for the ML surrogate models. In Figure~\ref{fig:KF_statistics}\textbf{c}, the temporal autocorrelation $R_{uu}(t^*)$ (See SI S6) confirms that both LR-AR and Uni-Flow sustain long-term dynamical consistency, matching the decorrelation behavior of the reference data.
The time-averaged kinetic energy spectra $\langle E(k)\rangle$  in Figure~\ref{fig:KF_statistics}\textbf{d} show that Uni-Flow preserves the multiscale energy cascade, unlike LR-AR predictions, which exhibit overestimated spectral damping at high wavenumbers. Quantitatively, Uni-Flow maintains close agreement with the reference spectrum, achieving a spectrum match up to two orders of magnitude lower than the LR-AR model in the high-wavenumber range ($k > 30$). Finally, the quantile-quantile plot of vorticity in Figure~\ref{fig:KF_statistics}\textbf{e} confirms that the overall vorticity distribution is well preserved, with Uni-Flow maintaining a closer match to the reference distribution in the tails compared to the LR-AR model.

% \XX{inference time}
We also evaluated the Fourier Neural Operator and the U-shaped Neural Operator in an ablation study, both of which are unable to capture the high frequency spectrum for the $256\times256$ under a similar number of parameters. 
% Cross-reference to Supplementary Info
The Uni-Flow network and the total number of parameters of the Uni-Flow setup are presented in Supplementary information S2. The inference time against the numerical solver, LR-AR model, is demonstrated in Supplementary information S3. 
These results highlight Uni-Flow's ability to evolve interpretable, low-dimensional latent dynamics while reconstructing physically faithful and statistically consistent 2D turbulence, providing strong validation of the core framework.

\subsection{3D Turbulent channel flow inflow generator}
% Problem introduction
The turbulent channel flow represents a canonical wall-bounded shear flow characterised by strong velocity gradients and coherent near-wall structures. Accurate inflow generation for such systems is essential for large-eddy and direct numerical simulations~\cite{wu2017inflow,schluter2004large,poletto2011divergence}, where realistic temporal and spatial correlations must be maintained. 
% Mathematical formulation
In this application, we employ the lattice Boltzmann method (see Methods~C) to generate reference turbulent inflow data~\cite{xue2022synthetic}. To ensure high-fidelity turbulence statistics, a three-dimensional turbulent channel flow is simulated with periodic boundary conditions in the streamwise and spanwise directions. The simulation is performed at a friction Reynolds number of $Re_{\tau}=180$ in a computational domain of $2H \times 2H \times 6\pi H$, with $H=96$ grid points in the half-channel height, providing sufficient resolution to capture all near-wall structures. Each simulation is advanced for 1.5 million timesteps, and data are extracted once the flow reaches a statistically stationary state. The instantaneous velocity fields from the mid-plane cross-section are used as training samples for the inflow generator. The resulting dataset comprises 320 trajectories (320 temporal snapshots) each with spatial dimensions of $192\times192$ grid points. The data are divided into training, validation, and test subsets with an 80/10/10 split. Further details of the numerical setup and data preprocessing are provided in Supplementary Information Section S1.2.

\begin{figure}[htbp!]
\centering
\includegraphics[width=1\textwidth]{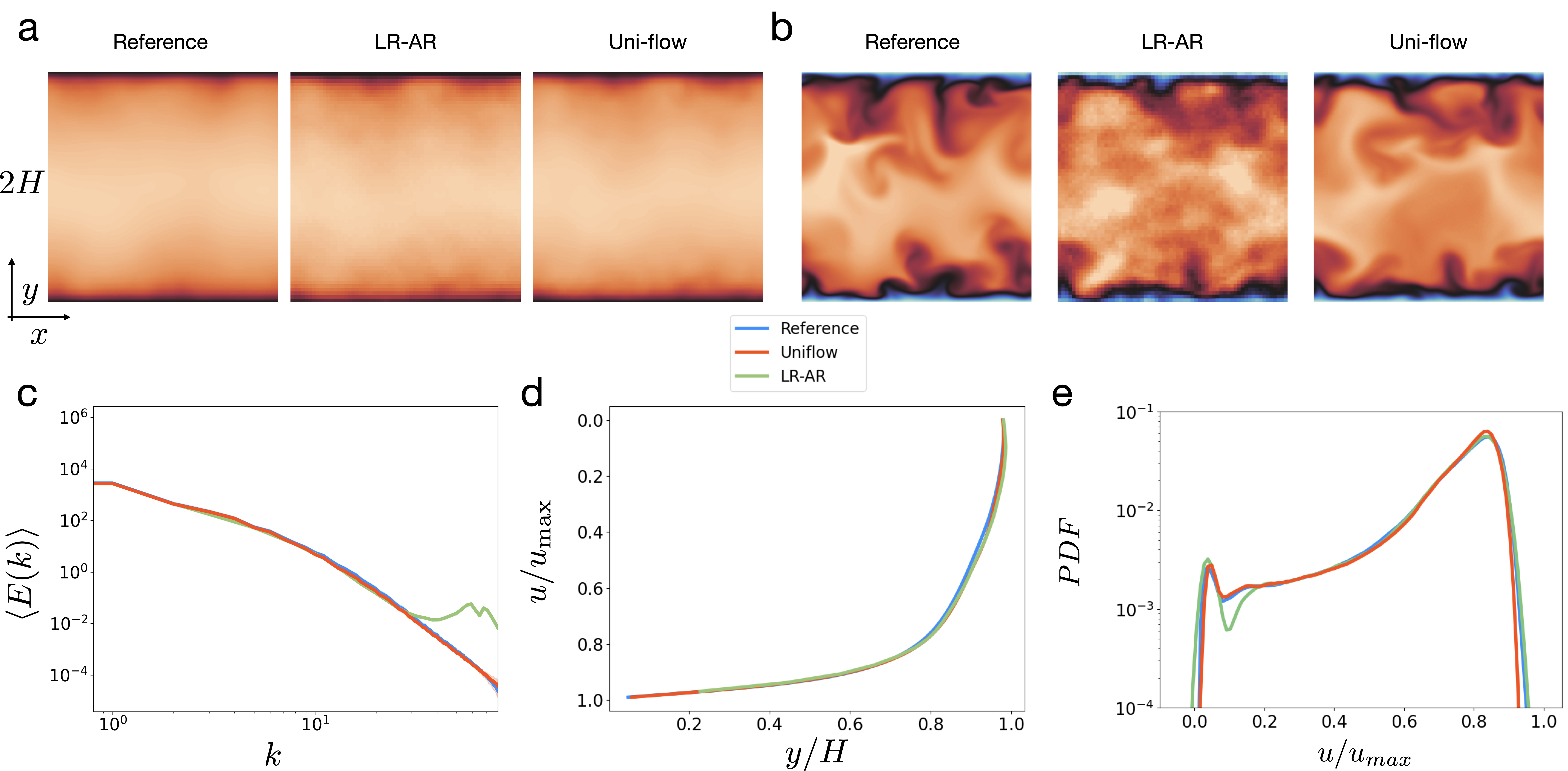}
%\label{fig:periodic-stg}
\caption{\textbf{Demonstration of Uni-Flow for generating turbulent inflow conditions in a channel-flow configuration.}  Panel (a) presents the time-averaged velocity comparison between Reference, LR-AR, and Uni-Flow over 600 time frames. Panel (b) shows instantaneous streamwise velocity fields for the reference LBM-LES (left), LR-AR baseline (middle), and Uni-Flow prediction (right).  Panel (c) displays the streamwise energy spectrum, confirming that Uni-Flow recovers both the inertial and dissipative ranges with higher fidelity than LR-AR. The blue, red and green lines are representing reference data, Uni-Flow, and LR-AR cases respectfully. Panel (d) shows the time-averaged normalised turbulence velocity profiles, where Uni-Flow matches the reference at all scales, whereas, LR-AR failed to match near the wall. Panel (e) presents the probability density function of normalised velocity distribution, demonstrating that Uni-Flow reproduces the correct velocity distribution whereas LR-AR case failed to capture the low velocity region. }
\label{fig:tcf_statistics}
\end{figure}

% Objective: State the learning task clearly: what is being learned.
The goal is to learn a generative model of the turbulent inflow that enables stable long-time rollouts while preserving the statistical properties of the reference flow at a high spatial resolution of $192 \times 192$. For the LR-AR component, Uni-Flow admits a broad class of operator-learning models in principle. However, for the turbulent inflow dataset considered here, we observe that conventional Koopman formulations are unable to provide stable and sufficiently expressive latent dynamics, leading to degraded long-horizon rollouts.
% To address this dataset-specific limitation, we employ a quantum-informed Koopman operator, which provides a stable and expressive latent representation for this particular inflow configuration, 
To ensure stable and expressive latent temporal dynamics for this dataset, we employ a quantum-informed Koopman operator with priors trained on a 20-qubit quantum device~\cite{wangxue2025quantum}. This choice showcases a pathway towards hybrid quantum-classical surrogates, although the Uni-Flow framework itself remains architecture-agnostic. The high-resolution fields are then reconstructed using DDIM sampling with 40 denoising steps.
% In the LR-AR component, a quantum-informed Koopman operator~\cite{wangxue2025quantum} is employed to guide the evolution of low-resolution spatiotemporal dynamics, while the high-resolution fields are reconstructed using DDIM sampling with 40 denoising steps.

% results
Figure~\ref{fig:tcf_statistics} demonstrates the long-term rollout of Uni-Flow to reconstruct physically consistent turbulent inflow conditions for channel flow at a friction Reynolds number $Re_{\tau}=180$. In all cases, the blue, red, and green lines represent the reference data, Uni-Flow, and LR-AR results, respectively. Figure~\ref{fig:tcf_statistics}\textbf{a} shows time averaged streamwise velocity fields for the LBM simulation reference (left), the LR-AR baseline (middle), and Uni-Flow (right). Both LR-AR and Uni-Flow are able to capture the long-term autoregressive rollout mean field. Figure~\ref{fig:tcf_statistics}\textbf{b} shows instantaneous snapshots of the reference, LR-AR, and Uni-Flow spatiotemporal fields. The streamwise velocity contours reveal that Uni-Flow accurately preserves both the large structures of the turbulent channel flow and the small velocity fluctuations in the near-wall region, whereas LR-AR fails to capture the structures observed in the reference simulation due to resolution and noise. The time-averaged energy spectrum in Figure~\ref{fig:tcf_statistics}\textbf{c} confirms that Uni-Flow accurately captures the energy across all scales of the turbulent channel flow. However, the LR-AR case is only able to capture the large scale structure and may not represent the small scale energy fluctuations. Figure~\ref{fig:tcf_statistics}\textbf{d} compares the time-averaged normalised turbulence velocity profiles as a function of  $y/H$. , showing close agreement between Uni-Flow and the reference, whereas LR-AR underestimates the near-wall peak.
Finally, Figure~\ref{fig:tcf_statistics}\textbf{e} shows the probability density function (PDF) of the normalised streamwise velocity, demonstrating that Uni-Flow recovers the correct distribution, indicating that the generated inflow fields reproduce the full statistical distribution of turbulent fluctuations. The LR-AR case failed to preserve the small velocity fluctuations.
% Cross-reference to Supplementary Info
Supplementary Information S2 represents the neural network architectures and parameters.
% Concise closing
These results demonstrate that Uni-Flow generates statistically consistent inflow fields that sustain long-term stability and reproduce the spectral and structural features of wall-bounded turbulence at high resolution. From this engineering problem, we turn to a time-sensitive biomedical application, demonstrating the framework's potential for real-time physiological modelling.
% To demonstrate the framework's adaptability to emerging hardware and algorithms, we specifically implement a quantum-informed Koopman operator, where priors are trained on a 20-qubit quantum device~\cite{wangxue2025quantum}. This choice showcases a pathway towards hybrid quantum-classical surrogates, although the Uni-Flow framework itself remains architecture-agnostic.

\subsection{Stenotic aortic flow}
% Problem introduction
To assess the applicability of the Uni-Flow framework to physiological flow modelling, we apply it to pulsatile blood flow through a patient-specific stenotic aorta reconstructed from imaging data~\cite{xue2025uncertainty,lo2025multi}. This case represents a complex, unsteady, and geometrically constrained flow environment characterised by transitional turbulence and strong wall-shear gradients, providing a stringent test for generative modelling under physiological conditions. The stenotic aortic flow serves as a representative biomedical example for assessing whether Uni-Flow can reproduce temporally resolved hemodynamics and spatial pressure distributions consistent with HemeLB simulations~\cite{mazzeo2008hemelb}. In particular, the model is expected to capture the cyclic flow acceleration and deceleration associated with cardiac pulsation while preserving the spatial coherence of high-shear regions near the stenotic throat.
% Mathematical formulation
High-fidelity reference data are generated using the HemeLB solver~\cite{mazzeo2008hemelb,xue2024lattice} (see Methods C).

\begin{figure}[htbp!]
\centering
\includegraphics[width=1\textwidth]{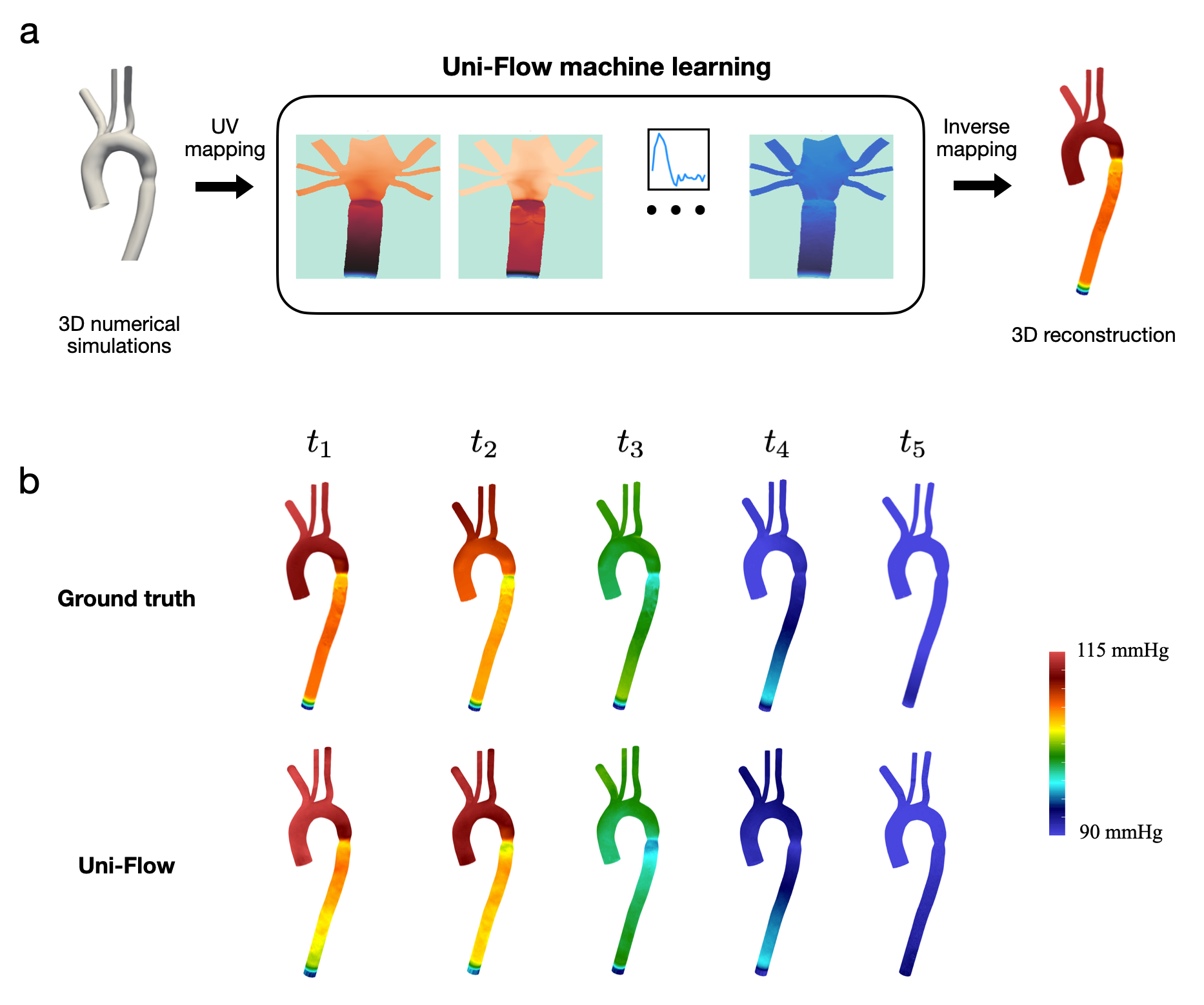}
%\label{fig:periodic-stg}
\caption{\textbf{Uni-Flow learning of spatiotemporal pressure dynamics in aortic stenosis.} Panel (a) Overview of the Uni-Flow framework for complex cardiovascular flows. Three-dimensional numerical simulations (HemeLB) of a stenotic aorta are mapped to a two-dimensional UV domain, where Uni-Flow learns the spatiotemporal evolution of wall pressure fields entirely in 2D. The predicted fields are then inversely mapped to the three-dimensional geometry to reconstruct the physical distribution on the aortic surface. Panel (b) Comparison of wall pressure evolution across a cardiac cycle at 5 representative time frames ($t_1$ to $t_5$). Uni-Flow accurately reproduces the high-pressure buildup upstream of the stenosis and the downstream recovery region, closely matching the ground-truth simulation results both spatially and temporally.}
\label{fig:coa}
\end{figure}
% Scenarios / cases
We use three-dimensional HemeLB simulation data of pulsatile flow in a patient-specific aortic geometry with a physiological inflow waveform and a peak centreline velocity of approximately $1~\mathrm{m\,s^{-1}}$, corresponding to a Reynolds number of $Re \approx 4600$ based on the inlet diameter and mean systolic velocity. The dataset spans 100 cardiac cycles and is divided into 80\% for training, 10\% for validation, and 10\% for testing. Training sequences are extracted from the statistically periodic regime after initial transients (4 cardiac cycles). The detailed information regarding the dataset generation is demonstrated in Supplementary Information S1.3. The physics-informed LR-AR module, driven by the norm-preserving Koopman operator~\cite{mezic2021koopman,brunton2021modern,wangxue2025quantum}, learns the latent temporal dynamics in the rollout phase, ensuring stable long-horizon evolution without gradient divergence. At each prediction step, the mask-based diffusion model refines the reconstructed fields to enhance spatial accuracy, focusing on high-gradient regions such as the stenotic jet and near-wall stress concentrations. This two-stage design allows Uni-Flow to first capture the global temporal evolution and then apply localised corrections to pressure and wall-shear stress distributions. 
% All surface quantities are parameterised onto a $512 \times 512$ UV map to facilitate structured learning, 
All surface quantities are mapped onto a $512 \times 512$ UV grid as a task-driven structured representation for learning, assuming that the resulting mapping preserves neighbourhood relations and spatial coherence relevant to the target observables. The learning task is formulated as a two-stage operator. The low-resolution autoregressive component $\mathcal{G}_\mathrm{LR}$ evolves the latent spatiotemporal dynamics of aortic pressure, while the high-resolution refinement operator $\mathcal{R}$ reconstructs detailed spatial fields of pressure $\{p^{\mathrm{SR}}_t\}$ from the evolving latent state. This hybrid formulation enables Uni-Flow to capture the global pulsatile dynamics at a low cost (with only 8.3 Million parameters, see Supplementary information S2) while recovering high-resolution hemodynamic quantities in the stenotic region. 

Figure~\ref{fig:coa}\textbf{a} illustrates the Uni-Flow framework for learning surface pressure dynamics in aortic stenosis. Three-dimensional HemeLB simulations of the stenotic aorta, driven by realistic heartbeat profiles~\cite{lo2024multi}, are first mapped onto a two-dimensional UV domain ($512\times512$ grid), where Uni-Flow learns the spatiotemporal evolution of wall pressure fields before inversely mapping the predictions back to the three-dimensional geometry to reconstruct the physical distributions. Figure~\ref{fig:coa}\textbf{b} compares the spatiotemporal evolution of wall pressure between Uni-Flow predictions and the reference HemeLB simulation. The predicted pressure fields closely reproduce the ground-truth dynamics throughout the cardiac cycle, accurately capturing the propagation of the systolic pressure wave, the peak pressure drop across the stenotic throat, and the subsequent recovery during diastole. The smooth transition between high- and low-pressure regions indicates that Uni-Flow preserves spatial continuity and avoids numerical artifacts or phase lag over multiple cycles. Quantitatively, the predicted pressure amplitude and gradient distribution fall within the expected range for the simulated physiological conditions of 90-115 mmHg, consistent with the simulation reference. These results confirm that the Uni-Flow framework reconstructs temporally coherent and spatially resolved pressure evolution under realistic hemodynamic conditions in aortic stenosis.

\begin{figure}[htbp!]
\centering
\includegraphics[width=0.9\textwidth]{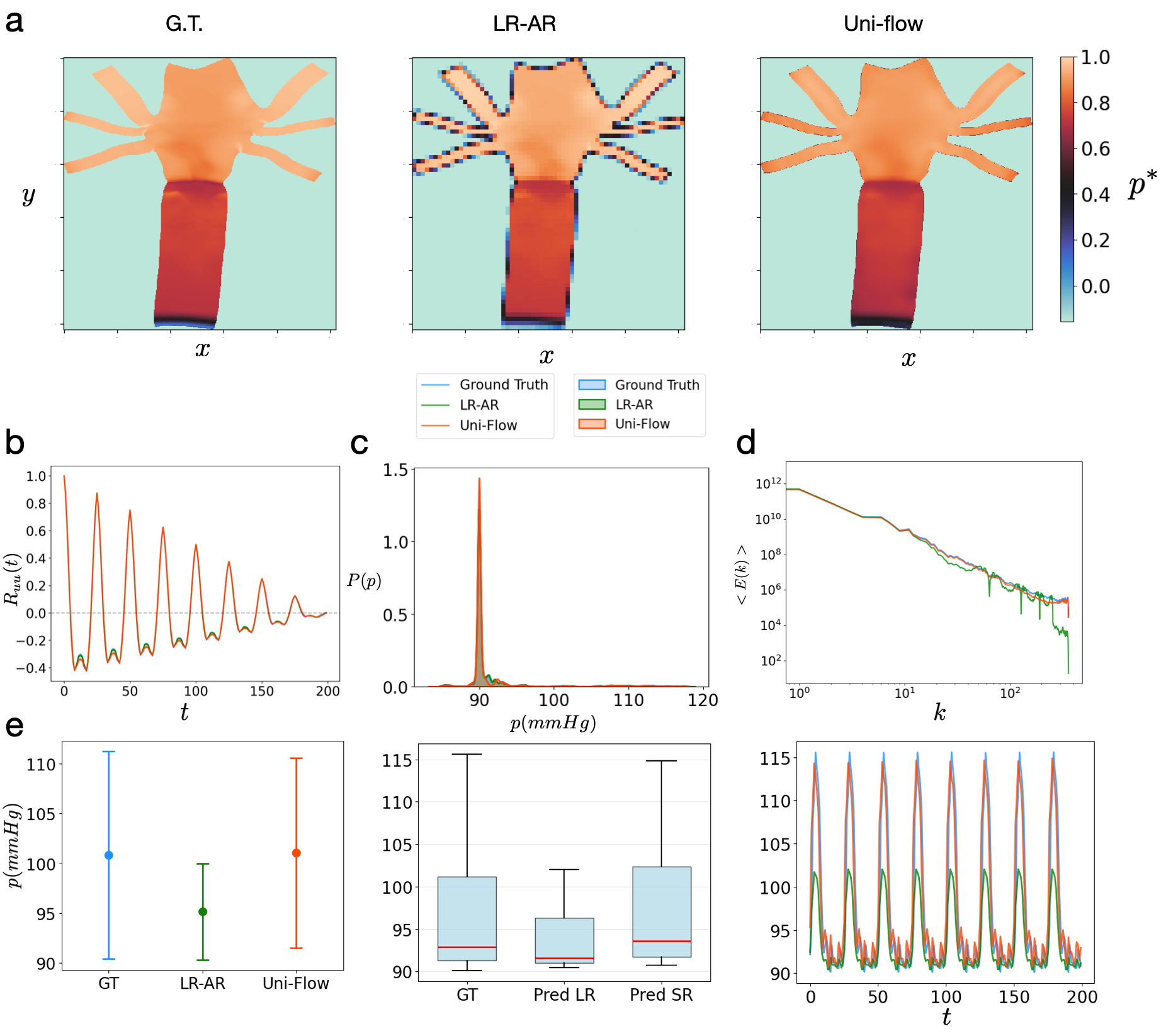}
%\label{fig:periodic-stg}
\caption{\textbf{ Quantitative assessment of pressure dynamics in stenotic aortic flow.} Panel (a) shows the ensamble average surface pressure distributions on the aortic wall for the ground truth (left), LR-AR (middle), and Uni-Flow (right), where Uni-Flow accurately captures the high-pressure region upstream of the stenosis and the recovery downstream. Panel (b) presents the temporal autocorrelation functions, showing that Uni-Flow maintains periodic coherence across cardiac cycles. Panel (c) displays the pressure kernel density estimate plots, indicating that Uni-Flow reproduces the sharp, narrow distribution observed in the HemeLB reference. Panel (d) shows the pressure energy spectra for the peak of heart beat cycle, demonstrating that Uni-Flow recovers the correct spectral scaling and dissipation range across all resolved wavenumbers, while LR-AR model failed to capture the high wave number energy spectrum. Panel (e) quantifies the pressure-drop characteristics within the highlighted region over 8 cardiac cycles (200 inference frames), comparing peak amplitudes, statistical distributions, and temporal waveforms. Uni-Flow remains within the simulated physiological range of 90-115~mmHg, accurately reproducing pulsatile amplitude and phase while matching the mean and variability of the reference. In contrast, the LR-AR baseline underestimates systolic peaks and shows reduced dynamic range. }
\label{fig:coa_stats}
\end{figure}

Figure~\ref{fig:coa_stats} provides a quantitative comparison of the reference simulation, the LR-AR component, and Uni-Flow results for the stenotic aortic flow. Figure~\ref{fig:coa_stats}\textbf{a} compares the time-averaged surface pressure distributions over 20 cardiac cycles on the aortic wall for the ground truth from HemeLB (left), the LR-AR model (middle), and Uni-Flow (right). Uni-Flow accurately reconstructs the spatial pressure gradients, particularly within the stenotic region, capturing both the upstream high-pressure buildup and the downstream pressure drop with good agreement. In Figure~\ref{fig:coa_stats}\textbf{b}, the temporal autocorrelation of the pressure signal demonstrates that Uni-Flow maintains coherent periodicity and stable phase alignment across cardiac cycles, whereas LR-AR exhibits gradual phase drift. Figure~\ref{fig:coa_stats}\textbf{c} presents the pressure probability density functions, showing that both Uni-Flow and LR-AR reproduce the sharp, narrow distribution of rollout predictions over 20 cardiac cycles, which matches the ground truth data. The pressure energy spectra in Figure~\ref{fig:coa_stats}\textbf{d} further reveal that Uni-Flow recovers the correct spectral scaling and dissipation range across all resolved wavenumbers, outperforming the LR-AR in properly capturing the high wavenumber region. Figure~\ref{fig:coa_stats}\textbf{e} quantifies the pressure-drop behaviour within the highlighted region over 8 cardiac cycles (200 inference frames), including peak amplitudes, distributions, and temporal waveforms. Uni-Flow remains within the physiological range of 90-115~mmHg, accurately reproducing the pulsatile amplitude and phase while matching the mean and variability of the HemeLB reference. In contrast, LR-AR underestimates systolic peaks and exhibits a reduced dynamic range. These analyses show that the LR-AR component effectively captures the underlying temporal dynamics of the stenotic flow. Uni-Flow further reconstructs high-resolution spatial features in strong agreement with the reference simulation, achieving a high-resolution representation of complex aortic hemodynamics.
% \FloatBarrier
\vspace{0em}
% Cross-reference to Supplementary Info
Supplementary Section~S2 summarises the neural network architecture and parameter configuration. 
% In the stenotic aortic flow case, Uni-Flow enables faster than real-time inference of physiologically relevant hemodynamic evolution, reducing a multi-hour high-fidelity simulation to second-scale prediction on a single GPU. 
In the stenotic aortic flow case, Uni-Flow is formulated to infer physiologically relevant surface pressure dynamics directly, rather than to reproduce the full three-dimensional high-fidelity flow simulation.
% Specifically, a 40-second pulsatile hemodynamic evolution that requires 8.28 hours of computation on 128 GPUs using the HemeLB solver is inferred in 27.5 seconds using Uni-Flow, corresponding to an approximate $1.4\times10^5$ task-level
% wall-clock speedup for generating surface pressure time series (see Supplementary Information Section S3).
By evolving low-resolution latent temporal dynamics and applying diffusion-based refinement solely to recover spatial detail consistent with this constraint, the model operates in a reduced spatiotemporal subspace aligned with the target observables.
As a consequence of this structural decomposition, a 40-second pulsatile haemodynamic evolution that requires 8.28 hours of computation on 128 GPUs using the HemeLB solver can be inferred in 27.5 seconds on a single GPU using Uni-Flow, corresponding to an approximate $1.4\times10^5$ task-level wall-clock speedup for generating surface pressure time series (see Supplementary Information Section~S3).

\section{Conclusion}\label{sec:conclusion}

We have introduced Uni-Flow, a unified autoregressive–diffusion framework for generative modelling of complex multiscale flow dynamics. By coupling a low-dimensional autoregressive component that governs long-horizon temporal evolution with a diffusion-based refinement process for spatial reconstruction, Uni-Flow decouples temporal stability from spatial fidelity. As a result, across all cases, Uni-Flow accurately reproduces instantaneous flow fields, long-term dynamical behaviour, and time-invariant statistical properties, including energy spectra, correlation functions, and probability distributions, while maintaining high spatial fidelity. In the hemodynamics setting, Uni-Flow enables faster-than-real-time inference of pulsatile hemodynamic evolution while preserving key pressure and flow characteristics across stenotic regions. This computational capability demonstrates the utility of generative surrogate models for time-sensitive hemodynamic analysis and exploratory modelling. Beyond its current instantiations with Fourier- and Koopman-based autoregressive modules, Uni-Flow is inherently architecture-agnostic at the low-resolution temporal modelling level, allowing alternative operator-learning or neural sequence models to be readily incorporated. More broadly, Uni-Flow establishes a connection between generative flow modelling, quantum machine learning, and scientific computing, opening new possibilities for hybrid quantum–classical surrogates and scalable, physics-grounded modelling of complex multiscale systems. These results position Uni-Flow as a general computational framework for advancing real-time and long-horizon simulation across physics, biology, and engineering.

\section{Methods}
\subsection{Low-resolution autoregressive models}
This subsection describes the low-resolution autoregressive  component of Uni-Flow, which is responsible for learning stable long-horizon temporal dynamics in a reduced spatial representation. Uni-Flow admits a broad class of operator-learning models for the LR-AR component; the specific architectures described below are chosen based on dataset characteristics rather than methodological necessity.

\subsubsection{Fourier Neural Operator}

In the KS-2D and two-dimensional Kolmogorov flow systems, the low-resolution temporal evolution is learned using the Fourier Neural Operator (FNO) in an autoregressive rollout manner~\cite{li2020fourier,li2021learning}. We consider discrete-time dynamics
\begin{equation}
\mathbf{u}_{n+1} = \mathcal{G}(\mathbf{u}_{n}),
\end{equation}
where $\mathbf{u}_{n}$ denotes the spatial field at time step $t_n$. The FNO approximates the nonlinear operator $\mathcal{G}$ via spectral convolution in Fourier space.

Let $\mathcal{F}$ and $\mathcal{F}^{-1}$ denote the discrete Fourier transform and its inverse. 
Define $\hat{\mathbf{u}}_{n}(\mathbf{k}) = \mathcal{F}[\mathbf{u}_{n}](\mathbf{k})$, where $\mathbf{k}$ is the discrete wavenumber index. Each FNO layer performs a spectral transformation restricted to a truncated set of modes $|\mathbf{k}| \le k_{\max}$:

\begin{equation}
\hat{\mathbf{v}}(\mathbf{k}) 
= 
\mathbf{W}(\mathbf{k}) \, \hat{\mathbf{u}}_{n}(\mathbf{k}) 
+ 
\mathbf{b}(\mathbf{k}), 
\quad |\mathbf{k}| \le k_{\max},
\end{equation}
with $\hat{\mathbf{v}}(\mathbf{k}) = 0$ for higher modes. 
Here $\mathbf{W}(\mathbf{k})$ are complex-valued learnable weights and $\mathbf{b}(\mathbf{k})$ denotes a spectral bias term. 
The result is mapped back to physical space through an inverse Fourier transform and combined with a residual linear projection:
\begin{equation}
\mathbf{u}_{n+1} 
=
\sigma\!\left(
\mathcal{F}^{-1}[\hat{\mathbf{v}}(\mathbf{k})]
+
\mathbf{W}_r \mathbf{u}_{n}
\right),
\end{equation}
where $\sigma(\cdot)$ denotes the GELU activation and $\mathbf{W}_r$ is a learned pointwise linear operator that preserves low-frequency information through a residual connection. FNO provides a resolution-invariant mapping between function spaces, enabling the efficient learning of spatiotemporal dynamics governed by partial differential equations. Nonetheless, as demonstrated in Supplementary Information Section S5, both the FNO and U-shaped Neural Operator exhibit difficulties in capturing high-wavenumber components, which limits their ability to resolve fine-scale structures. Accordingly, for the first two systems, we employ FNO primarily to model low-resolution dynamics constructed from low-wavenumber components.

For the two-dimensional Kolmogorov flow, the FNO is trained in a supervised autoregressive manner to predict the next-step low-resolution vorticity field. 
Given $\omega_n^{\mathrm{LR}} \in \mathbb{R}^{64 \times 64}$, the network learns the mapping
\[
\omega_{n+1}^{\mathrm{LR}} 
= 
\mathcal{G}_{\mathrm{FNO}}(\omega_n^{\mathrm{LR}}).
\]
The training objective is a variance-normalised root-mean-square error (vRMSE),
\begin{equation}
\mathcal{L}_{\mathrm{NO}}
=
\frac{
\sqrt{
\frac{1}{|\mathcal{D}_m|}
\left\|
\mathbf{u}_{n+1}
-
\mathcal{G}_{\mathrm{FNO}}(\mathbf{u}_{n})
\right\|_2^2
}
}{
\sigma(\mathbf{u}_{n+1})
},
\end{equation}
where $|\mathcal{D}_m|$ denotes the number of spatial grid points (domain normalisation), and $\sigma(\mathbf{u}_{n+1})$ is the spatial standard deviation of the reference field.

\subsubsection{Koopman-based latent linear model}

For turbulent inflow generation and aortic stenosis, we approximate $\mathcal{G}$ using a Koopman-based latent linear dynamics model. An encoder-decoder pair $(\phi,\psi)$ lifts the low-resolution state $\mathbf{u}_n$ to $\mathbf{z}_n\in\mathbb{R}^d$, and the latent evolution is linear:

\begin{equation}
\mathbf{z}_{n}=\phi(\mathbf{u}_{n}),\quad
\mathbf{z}_{n+1}=\mathbf{K}\mathbf{z}_{n},\quad
\hat{\mathbf{u}}_{n}=\psi(\mathbf{z}_{n}).
\end{equation}
where $\mathbf{z}_{n}\in\mathbb{R}^{d}$ and $\mathbf{K}\in\mathbb{R}^{d\times d}$ is the learned latent transition operator (a finite-dimensional approximation of the Koopman action). To stabilise long-horizon rollouts, we regularise $\mathbf{K}$ to be approximately norm-preserving.  In the real-valued implementation used here, this corresponds to an orthogonality constraint $\mathbf{K}^\top \mathbf{K}\approx \mathbf{I}$ (equivalently, unitarity $\mathbf{K}^\ast\mathbf{K}=\mathbf{I}$ in the complex-valued case). This penalty measures the orthogonality defect of $\mathbf{K}$ and encourages its columns to be approximately orthonormal. We implement this using the Frobenius-norm penalty
\begin{equation}
\mathcal{L}_{\mathrm{ortho}} = \|\mathbf{K}^\top \mathbf{K} - \mathbf{I}\|_F^2,
\end{equation}
which mitigates artificial growth or decay of the latent-state norm during autoregressive evolution~\cite{wangxue2025quantum}. The total training loss function is
\begin{equation}
\mathcal{L}_{\mathrm{total}} =
\mathcal{L}_{\mathrm{recon}} + \mathcal{L}_{\mathrm{pred}}
+ \lambda_{\mathrm{ortho}} \mathcal{L}_{\mathrm{ortho}}
+ \lambda_{\mathrm{KL}} \mathcal{L}_{\mathrm{KL}}
+ \lambda_{\mathrm{MMD}} \mathcal{L}_{\mathrm{MMD}}.
\end{equation}
The reconstruction loss enforces autoencoder consistency,
\begin{equation}
\mathcal{L}_{\mathrm{recon}} =
\|\psi(\phi(\mathbf{u}_{n}))-\mathbf{u}_{n}\|_2^2
+
\|\psi(\phi(\mathbf{u}_{n+1}))-\mathbf{u}_{n+1}\|_2^2,
\end{equation}
and the prediction loss enforces accurate one-step evolution in the physical space,
\begin{equation}
\mathcal{L}_{\mathrm{pred}} =
\|\psi(\mathbf{K}\phi(\mathbf{u}_{n}))-\mathbf{u}_{n+1}\|_2^2
+
\|\psi(\mathbf{K}^\top\phi(\mathbf{u}_{n+1}))-\mathbf{u}_{n}\|_2^2,
\end{equation}
where the second term provides backward-consistency; when $\mathbf{K}$ is approximately orthogonal, $\mathbf{K}^\top$ approximates $\mathbf{K}^{-1}$. The statistical alignment terms are applied prior to encoding, on a fixed observable map $\eta(\mathbf{u})$ of the low-resolution state. 
We match $\hat{q}(\eta(\mathbf{u}))$ to the quantum-informed prior $p_{\theta}(\eta(\mathbf{u}))$ using $\mathcal{L}_{\mathrm{KL}}$ and $\mathcal{L}_{\mathrm{MMD}}$. For the aortic stenosis case, we set $\lambda_{\mathrm{KL}}=\lambda_{\mathrm{MMD}}=0$ since no quantum-informed prior is used, whereas for the turbulent inflow case we set $\lambda_{\mathrm{KL}}=\lambda_{\mathrm{MMD}}=1$.

\begin{algorithm}[H]
\caption{Training a Denoising Diffusion Probabilistic Model}
\label{alg:ddpm_training}
\begin{algorithmic}[1]
\Require Dataset $\mathcal{D} = \{x_0\}$, noise schedule $\{\beta_s\}_{s=1}^T$, model $\hat{\varepsilon}_\theta(x_s, s)$
\State Compute $\alpha_s = 1 - \beta_s$, and cumulative product $\bar{\alpha}_s = \prod_{r=1}^s \alpha_r$
\For{number of training steps}
    \State $x_0 \sim \mathcal{D}$; $s \sim \text{Uniform}(\{1, \ldots, T\})$; $\varepsilon \sim \mathcal{N}(0, I)$
    \State $x_s = \sqrt{\bar{\alpha}_s} x_0 + \sqrt{1 - \bar{\alpha}_s}\,\varepsilon$
    \State $\hat{\varepsilon} = \hat{\varepsilon}_\theta(x_s, s)$
    \State $\mathcal{L}_\text{ddpm} = \|\varepsilon - \hat{\varepsilon}\|^2$
    \State Update model parameters by the optimizer via backpropagation
\EndFor
\State \Return Trained denoising model $\hat{\varepsilon}_\theta$
\end{algorithmic}
\end{algorithm}

\subsection{Diffusion model for small scale dynamics refinement}

Within Uni-Flow, diffusion models are deliberately restricted to spatial refinement and are not used to advance temporal dynamics, allowing temporal causality to be enforced solely by the low-resolution autoregressive component. 
This design choice allows temporal stability and causal consistency to be enforced by the autoregressive component, while diffusion is leveraged solely to enhance spatial fidelity. Accordingly, we adopt diffusion models as a generative refinement mechanism for physical fields. Specifically, the Diffusion Denoising Probabilistic Model (DDPM)~\cite{ho2020ddpm} defines a forward diffusion process in which samples from a data distribution are gradually corrupted by Gaussian noise, providing a principled framework for learning the inverse denoising dynamics. This process is expressed as
\begin{equation}
x_s = \sqrt{\bar{\alpha}_s}\,x_0 + \sqrt{1-\bar{\alpha}_s}\,\varepsilon,
\qquad \varepsilon \sim \mathcal{N}(0,\mathbf{I}),
\end{equation}
where $\bar{\alpha}_s=\prod_{r=1}^{s}\alpha_r$, $\alpha_s=1-\beta_s$, and $\{\beta_s\}_{s=1}^{T}$ denotes the noise schedule with $s\in\{0,\ldots,T\}$ and $T\in\mathbb{Z}^+$. 
As $s$ increases, the field becomes progressively noisier, and the clean sample is recovered at $s=0$. A denoising network $\hat{\varepsilon}_{\theta}(x_s,s)$ is trained to predict the injected noise $\varepsilon$, and is subsequently used to define the reverse-time denoising dynamics for sampling. 
The complete training procedure is summarised in Algorithm~\ref{alg:ddpm_training}.
\begin{algorithm}[H]
\caption{High-Resolution reconstruction via DDIM Sampling Steps}
\label{alg:ddim_sampling}
\begin{algorithmic}[1]
\Require Low-resolution field $y$, upsampler $U$, DDPM noise predictor $\hat{\varepsilon}_\theta$, schedule $\{\bar{\alpha}_s\}_{s=0}^T$; steps $0=\tau_0<\cdots<\tau_K=T$
\State \textbf{Initialize (High-resolution guess):} $x \leftarrow U(y)$
\State \textbf{DDIM High-resolution reconstruction} $(x_{\tau_K} \rightarrow x_{\tau_0})$
\For{$k = K$ \textbf{to} $1$}
    \State $s \leftarrow \tau_k$, $r \leftarrow \tau_{k-1}$
    \State $\varepsilon_{\mathrm{pred}} \leftarrow \hat{\varepsilon}_\theta(x, s)$
    \State $\tilde{x}_0 \leftarrow \dfrac{x - \sqrt{1 - \bar{\alpha}_s}\,\varepsilon_{\mathrm{pred}}}{\sqrt{\bar{\alpha}_s}}$
    \State $x \leftarrow \sqrt{\bar{\alpha}_r}\,\tilde{x}_0 + \sqrt{1 - \bar{\alpha}_r}\,\varepsilon_{\mathrm{pred}}$
\EndFor
\State \Return $\hat{x}_0 \leftarrow x$
\end{algorithmic}
\end{algorithm}

% \begin{algorithm}[H]
% \caption{High-Resolution reconstruction via DDIM Sampling Steps}
% \label{alg:ddim_sampling}
% \begin{algorithmic}[1]
% \Require Low-resolution field $y$, upsampler $U$, DDPM noise predictor $\epsilon_\theta$, schedule $\{\bar\alpha_t\}_{t=0}^{T}$, steps $0=\tau_0<\cdots<\tau_K=T$

% \State \textbf{Initialize (High-resolution guess):} $x \gets U(y)$ 

% \State \textbf{DDIM High-resolution reconstruction} ($x_{\tau_K}\!\to\! x_{\tau_0}$)
% \For{$k=K$ \textbf{to} $1$}
%   \State $s \gets \tau_{k}$,\; $t \gets \tau_{k-1}$
%   \State $\epsilon \gets \epsilon_\theta(x,s)$
%   \State $\tilde{x}_0 \gets \dfrac{x - \sqrt{1-\bar\alpha_s}\,\epsilon}{\sqrt{\bar\alpha_s}}$
%   \State $x \gets \sqrt{\bar\alpha_t}\,\tilde{x}_0 + \sqrt{1-\bar\alpha_t}\,\epsilon$ \Comment{$x \equiv x_t$}
% \EndFor

% \State \Return $\hat{x}_0 \gets x$ \Comment{$x$ is back at $t=\tau_0=0$}
% \end{algorithmic}
% \end{algorithm}

% In the sampling phase, the network is used to transform a random noise sample into a meaningful physical field. This is achieved through the denoising step, given by
% \begin{equation}
% x_{s-1} = \frac{1}{\sqrt{\alpha_s}} \left( x_s - \frac{\beta_s}{\sqrt{1 - \bar{\alpha}_s}}  \hat{\varepsilon}_\theta(x_s, s) \right) + \xi\sqrt{\frac{1 - \bar{\alpha}_{s-1}}{1 - \bar{\alpha}_{s}}\beta_s}, \quad \xi \sim \mathcal{N}(0, I).
% \end{equation}

The original DDPM sampling procedure uses $T$ denoising iterations, one per discrete timestep in the diffusion schedule used during training~\cite{ho2020ddpm}, typically on the order of hundreds to thousands of steps. For super-resolution tasks based on diffusion inversion, as in our framework, this can be implemented by corrupting the low-resolution field with Gaussian noise corresponding to a chosen timestep $s$ (i.e., using the forward-process noise level determined by $\bar{\alpha}_s$), and using this corrupted sample as an approximation of the intermediate state~\cite{shu2023physics,chihaoui2024bird,wang2023sinsr}. The timestep (or equivalently the noise level) is treated as a tunable hyperparameter to maximize restoration quality.

Since DDPM sampling requires the same number of steps as used in training, employing the deterministic Denoising Diffusion Implicit Model (DDIM)~\cite{song2022ddim} can significantly improve efficiency~\cite{shu2023physics}, as it allows for the skipping of denoising steps while maintaining reconstruction quality. % The DDIM sampling procedure for high-resolution is summarised in Algorithm~\ref{alg:ddim_sampling}, where the initial estimate is obtained by upscaling the low-resolution field via pixel replication.
The DDIM sampling procedure for high-resolution reconstruction is summarised in Algorithm~\ref{alg:ddim_sampling}. The initial estimate is obtained by upscaling the low-resolution field via pixel replication, thereby fixing the coarse-scale dynamics with the low-resolution autoregressive model and leaving unresolved fine-scale variability to be reconstructed by diffusion. Here, $\{\tau_k\}_{k=0}^K$ denotes a subsampled version of the original DDPM timesteps $\{s\}_{s=0}^T$, allowing efficient deterministic sampling with fewer steps.

\subsection{Lattice Boltzmann method as the numerical approach to solve hemodynamics and turbulent flows}
Within this study, we make use of HemeLB~\cite{zacharoudiou2023development}, which is specifically optimised for sparse cardiovascular geometries and parallel execution. Its communication patterns and domain decomposition strategy are designed to minimise overhead in highly sparse domains, enabling excellent strong and weak scaling across modern supercomputers, including exascale platforms such as Frontier. We employ a three-dimensional lattice Boltzmann model using 19 discrete velocity directions, known as the D3Q19 model. Each lattice cell is defined by its spatial position $\mathbf{x}$ and time $t$, and is associated with a discrete velocity set $\mathbf{c}_i$ where $i \in \{0, 1, \ldots, Q-1\}$ and $Q = 19$. The evolution of the distribution functions follows the lattice Boltzmann equation:
\begin{equation}
\label{eq:lbe}
f_i(\mathbf{x}+\mathbf{c}_i \Delta t, t+\Delta t) = f_i(\mathbf{x}, t) - \Omega \left[f_i(\mathbf{x}, t) - f_i^{eq}(\mathbf{x}, t)\right],
\end{equation}
where $\Omega = \frac{\Delta t}{\tau}$ is the Bhatnagar-Gross-Krook (BGK) single-relaxation-time collision operator~\cite{succi2001lattice}, and $\tau$ is the relaxation time. The equilibrium distribution function $f_i^{eq}$ is defined as:
\begin{equation}
\label{eq:local_eq}
f_i^{eq}(\mathbf{x}, t) = w_i \rho(\mathbf{x}, t) \left[1 + \frac{\mathbf{c}_i \cdot \mathbf{u}(\mathbf{x}, t)}{c_s^2} + \frac{(\mathbf{c}_i \cdot \mathbf{u}(\mathbf{x}, t))^2}{2c_s^4} - \frac{\mathbf{u}(\mathbf{x}, t) \cdot \mathbf{u}(\mathbf{x}, t)}{2c_s^2} \right],
\end{equation}
where $w_i$ are the lattice weights ($w_0 = 1/3$, $w_{1-6} = 1/18$, $w_{7-18} = 1/36$), $\rho(\mathbf{x}, t)$ is the fluid density, and $\mathbf{u}(\mathbf{x}, t)$ is the macroscopic velocity. The time step $\Delta t$ is set to unity in lattice units. The kinematic viscosity $\nu$ is given by:
\begin{equation}
\label{eq:nu}
\nu = c_s^2 \left(\tau - \frac{1}{2}\right) \Delta t,
\end{equation}
where the lattice speed of sound satisfies $c_s^2 = 1/3$ in lattice units.

Macroscopic quantities are obtained from the moments of the distribution functions:
\begin{equation}
\label{eq:density}
\rho(\mathbf{x}, t) = \sum_{i=0}^{Q-1} f_i(\mathbf{x}, t),
\end{equation}
\begin{equation}
\label{eq:momentum}
\rho(\mathbf{x}, t)\mathbf{u}(\mathbf{x}, t) = \sum_{i=0}^{Q-1} f_i(\mathbf{x}, t)\mathbf{c}_i.
\end{equation}
Subgrid-scale turbulence modelling and outlet stabilisation are handled using standard Smagorinsky closures and sponge-layer techniques~\cite{smagorinsky1963general,koda2015lattice,xue2022synthetic}; full details are provided in the Supplementary Information Section S1.1. 

% \section{Data availability}
% The datasets used in this study are publicly available on Figshare:

% \begin{itemize}
%     \item Kolmogorov flow dataset: \url{https://figshare.com/kf_2d_256}
%     \item Turbulent channel inflow dataset: \url{https://figshare.com/tcf_192}
%     \item Stenotic aortic flow dataset: \url{https://figshare.com/coa_512}
% \end{itemize}

% \section{Code availability}
% The training and inference code for Uni-Flow is publicly available at the University College London Centre for Computational Science GitHub repository: \url{https://github.com/UCL-CCS/Uni-Flow}.

\bibliography{prex}

%aipnum4-2.bst 2019-01-14 (MD) hand-edited version of apsrev4-1.bst
\hyphenation{Post-Script Sprin-ger}
\begin{thebibliography}{65}%
\makeatletter
\providecommand \@ifxundefined [1]{%
 \@ifx{#1\undefined}
}%
\providecommand \@ifnum [1]{%
 \ifnum #1\expandafter \@firstoftwo
 \else \expandafter \@secondoftwo
 \fi
}%
\providecommand \@ifx [1]{%
 \ifx #1\expandafter \@firstoftwo
 \else \expandafter \@secondoftwo
 \fi
}%
\providecommand \natexlab [1]{#1}%
\providecommand \enquote  [1]{``#1''}%
\providecommand \bibnamefont  [1]{#1}%
\providecommand \bibfnamefont [1]{#1}%
\providecommand \citenamefont [1]{#1}%
\providecommand \href@noop [0]{\@secondoftwo}%
\providecommand \href [0]{\begingroup \@sanitize@url \@href}%
\providecommand \@href[1]{\@@startlink{#1}\@@href}%
\providecommand \@@href[1]{\endgroup#1\@@endlink}%
\providecommand \@sanitize@url [0]{\catcode `\\12\catcode `\$12\catcode
  `\&12\catcode `\#12\catcode `\^12\catcode `\_12\catcode `\%12\relax}%
\providecommand \@@startlink[1]{}%
\providecommand \@@endlink[0]{}%
\providecommand \url  [0]{\begingroup\@sanitize@url \@url }%
\providecommand \@url [1]{\endgroup\@href {#1}{\urlprefix }}%
\providecommand \urlprefix  [0]{URL }%
\providecommand \Eprint [0]{\href }%
\providecommand \doibase [0]{https://doi.org/}%
\providecommand \selectlanguage [0]{\@gobble}%
\providecommand \bibinfo  [0]{\@secondoftwo}%
\providecommand \bibfield  [0]{\@secondoftwo}%
\providecommand \translation [1]{[#1]}%
\providecommand \BibitemOpen [0]{}%
\providecommand \bibitemStop [0]{}%
\providecommand \bibitemNoStop [0]{.\EOS\space}%
\providecommand \EOS [0]{\spacefactor3000\relax}%
\providecommand \BibitemShut  [1]{\csname bibitem#1\endcsname}%
\let\auto@bib@innerbib\@empty
%</preamble>
\bibitem [{\citenamefont {Mazzeo}\ and\ \citenamefont
  {Coveney}(2008)}]{mazzeo2008hemelb}%
  \BibitemOpen
  \bibfield  {author} {\bibinfo {author} {\bibfnamefont {M.~D.}\ \bibnamefont
  {Mazzeo}}\ and\ \bibinfo {author} {\bibfnamefont {P.~V.}\ \bibnamefont
  {Coveney}},\ }\bibfield  {title} {\enquote {\bibinfo {title} {{HemeLB: A high
  performance parallel lattice-{B}oltzmann code for large scale fluid flow in
  complex geometries}},}\ }\href@noop {} {\bibfield  {journal} {\bibinfo
  {journal} {Computer Physics Communications}\ }\textbf {\bibinfo {volume}
  {178}},\ \bibinfo {pages} {894--914} (\bibinfo {year} {2008})}\BibitemShut
  {NoStop}%
\bibitem [{\citenamefont {Coveney}\ and\ \citenamefont
  {Wan}(2025)}]{coveney2025molecular}%
  \BibitemOpen
  \bibfield  {author} {\bibinfo {author} {\bibfnamefont {P.~V.}\ \bibnamefont
  {Coveney}}\ and\ \bibinfo {author} {\bibfnamefont {S.}~\bibnamefont {Wan}},\
  }\href@noop {} {\emph {\bibinfo {title} {{Molecular Dynamics: Probability and
  Uncertainty}}}}\ (\bibinfo  {publisher} {Oxford University Press},\ \bibinfo
  {year} {2025})\BibitemShut {NoStop}%
\bibitem [{\citenamefont {Evans}(2022)}]{evans2022partial}%
  \BibitemOpen
  \bibfield  {author} {\bibinfo {author} {\bibfnamefont {L.~C.}\ \bibnamefont
  {Evans}},\ }\href@noop {} {\emph {\bibinfo {title} {Partial differential
  equations}}},\ Vol.~\bibinfo {volume} {19}\ (\bibinfo  {publisher} {American
  mathematical society},\ \bibinfo {year} {2022})\BibitemShut {NoStop}%
\bibitem [{\citenamefont {Strikwerda}(2004)}]{strikwerda2004finite}%
  \BibitemOpen
  \bibfield  {author} {\bibinfo {author} {\bibfnamefont {J.~C.}\ \bibnamefont
  {Strikwerda}},\ }\href@noop {} {\emph {\bibinfo {title} {Finite difference
  schemes and partial differential equations}}}\ (\bibinfo  {publisher}
  {SIAM},\ \bibinfo {year} {2004})\BibitemShut {NoStop}%
\bibitem [{\citenamefont {Eymard}, \citenamefont {Gallou{\"e}t},\ and\
  \citenamefont {Herbin}(2000)}]{eymard2000finite}%
  \BibitemOpen
  \bibfield  {author} {\bibinfo {author} {\bibfnamefont {R.}~\bibnamefont
  {Eymard}}, \bibinfo {author} {\bibfnamefont {T.}~\bibnamefont
  {Gallou{\"e}t}},\ and\ \bibinfo {author} {\bibfnamefont {R.}~\bibnamefont
  {Herbin}},\ }\bibfield  {title} {\enquote {\bibinfo {title} {Finite volume
  methods},}\ }\href@noop {} {\bibfield  {journal} {\bibinfo  {journal}
  {Handbook of numerical analysis}\ }\textbf {\bibinfo {volume} {7}},\ \bibinfo
  {pages} {713--1018} (\bibinfo {year} {2000})}\BibitemShut {NoStop}%
\bibitem [{\citenamefont {Reddy}(1993)}]{reddy1993introduction}%
  \BibitemOpen
  \bibfield  {author} {\bibinfo {author} {\bibfnamefont {J.~N.}\ \bibnamefont
  {Reddy}},\ }\bibfield  {title} {\enquote {\bibinfo {title} {An introduction
  to the finite element method},}\ }\href@noop {} {\bibfield  {journal}
  {\bibinfo  {journal} {New York}\ }\textbf {\bibinfo {volume} {27}} (\bibinfo
  {year} {1993})}\BibitemShut {NoStop}%
\bibitem [{\citenamefont {Atchley}\ \emph {et~al.}(2023)\citenamefont
  {Atchley}, \citenamefont {Zimmer}, \citenamefont {Lange}, \citenamefont
  {Bernholdt}, \citenamefont {Melesse~Vergara}, \citenamefont {Beck},
  \citenamefont {Brim}, \citenamefont {Budiardja}, \citenamefont
  {Chandrasekaran}, \citenamefont {Eisenbach} \emph
  {et~al.}}]{atchley2023frontier}%
  \BibitemOpen
  \bibfield  {author} {\bibinfo {author} {\bibfnamefont {S.}~\bibnamefont
  {Atchley}}, \bibinfo {author} {\bibfnamefont {C.}~\bibnamefont {Zimmer}},
  \bibinfo {author} {\bibfnamefont {J.}~\bibnamefont {Lange}}, \bibinfo
  {author} {\bibfnamefont {D.}~\bibnamefont {Bernholdt}}, \bibinfo {author}
  {\bibfnamefont {V.}~\bibnamefont {Melesse~Vergara}}, \bibinfo {author}
  {\bibfnamefont {T.}~\bibnamefont {Beck}}, \bibinfo {author} {\bibfnamefont
  {M.}~\bibnamefont {Brim}}, \bibinfo {author} {\bibfnamefont {R.}~\bibnamefont
  {Budiardja}}, \bibinfo {author} {\bibfnamefont {S.}~\bibnamefont
  {Chandrasekaran}}, \bibinfo {author} {\bibfnamefont {M.}~\bibnamefont
  {Eisenbach}}, \emph {et~al.},\ }\bibfield  {title} {\enquote {\bibinfo
  {title} {Frontier: exploring exascale},}\ }in\ \href@noop {} {\emph {\bibinfo
  {booktitle} {Proceedings of the International Conference for High Performance
  Computing, Networking, Storage and Analysis}}}\ (\bibinfo {year} {2023})\
  pp.\ \bibinfo {pages} {1--16}\BibitemShut {NoStop}%
\bibitem [{\citenamefont {Hughes}(1995)}]{hughes1995multiscale}%
  \BibitemOpen
  \bibfield  {author} {\bibinfo {author} {\bibfnamefont {T.~J.}\ \bibnamefont
  {Hughes}},\ }\bibfield  {title} {\enquote {\bibinfo {title} {{Multiscale
  phenomena: Green's functions, the Dirichlet-to-Neumann formulation, subgrid
  scale models, bubbles and the origins of stabilized methods}},}\ }\href@noop
  {} {\bibfield  {journal} {\bibinfo  {journal} {Computer Methods in Applied
  Mechanics and Engineering}\ }\textbf {\bibinfo {volume} {127}},\ \bibinfo
  {pages} {387--401} (\bibinfo {year} {1995})}\BibitemShut {NoStop}%
\bibitem [{\citenamefont {Mason}(1994)}]{mason1994large}%
  \BibitemOpen
  \bibfield  {author} {\bibinfo {author} {\bibfnamefont {P.~J.}\ \bibnamefont
  {Mason}},\ }\bibfield  {title} {\enquote {\bibinfo {title} {Large-eddy
  simulation: A critical review of the technique},}\ }\href@noop {} {\bibfield
  {journal} {\bibinfo  {journal} {Quarterly Journal of the Royal Meteorological
  Society}\ }\textbf {\bibinfo {volume} {120}},\ \bibinfo {pages} {1--26}
  (\bibinfo {year} {1994})}\BibitemShut {NoStop}%
\bibitem [{\citenamefont {Piomelli}(1999)}]{piomelli1999large}%
  \BibitemOpen
  \bibfield  {author} {\bibinfo {author} {\bibfnamefont {U.}~\bibnamefont
  {Piomelli}},\ }\bibfield  {title} {\enquote {\bibinfo {title} {Large-eddy
  simulation: achievements and challenges},}\ }\href@noop {} {\bibfield
  {journal} {\bibinfo  {journal} {Progress in Aerospace Sciences}\ }\textbf
  {\bibinfo {volume} {35}},\ \bibinfo {pages} {335--362} (\bibinfo {year}
  {1999})}\BibitemShut {NoStop}%
\bibitem [{\citenamefont {Xue}, \citenamefont {Yao},\ and\ \citenamefont
  {Davidson}(2022)}]{xue2022synthetic}%
  \BibitemOpen
  \bibfield  {author} {\bibinfo {author} {\bibfnamefont {X.}~\bibnamefont
  {Xue}}, \bibinfo {author} {\bibfnamefont {H.-D.}\ \bibnamefont {Yao}},\ and\
  \bibinfo {author} {\bibfnamefont {L.}~\bibnamefont {Davidson}},\ }\bibfield
  {title} {\enquote {\bibinfo {title} {{Synthetic turbulence generator for
  lattice {B}oltzmann method at the interface between RANS and LES}},}\
  }\href@noop {} {\bibfield  {journal} {\bibinfo  {journal} {Physics of
  Fluids}\ }\textbf {\bibinfo {volume} {34}},\ \bibinfo {pages} {055118}
  (\bibinfo {year} {2022})}\BibitemShut {NoStop}%
\bibitem [{\citenamefont {Yang}\ and\ \citenamefont
  {Griffin}(2021)}]{yang2021grid}%
  \BibitemOpen
  \bibfield  {author} {\bibinfo {author} {\bibfnamefont {X.~I.}\ \bibnamefont
  {Yang}}\ and\ \bibinfo {author} {\bibfnamefont {K.~P.}\ \bibnamefont
  {Griffin}},\ }\bibfield  {title} {\enquote {\bibinfo {title} {Grid-point and
  time-step requirements for direct numerical simulation and large-eddy
  simulation},}\ }\href@noop {} {\bibfield  {journal} {\bibinfo  {journal}
  {Physics of Fluids}\ }\textbf {\bibinfo {volume} {33}},\ \bibinfo {pages}
  {015108} (\bibinfo {year} {2021})}\BibitemShut {NoStop}%
\bibitem [{\citenamefont {Shirvani}, \citenamefont {Nili-Ahmadabadi},\ and\
  \citenamefont {Ha}(2023)}]{shirvani2023machine}%
  \BibitemOpen
  \bibfield  {author} {\bibinfo {author} {\bibfnamefont {A.}~\bibnamefont
  {Shirvani}}, \bibinfo {author} {\bibfnamefont {M.}~\bibnamefont
  {Nili-Ahmadabadi}},\ and\ \bibinfo {author} {\bibfnamefont {M.~Y.}\
  \bibnamefont {Ha}},\ }\bibfield  {title} {\enquote {\bibinfo {title} {Machine
  learning-accelerated aerodynamic inverse design},}\ }\href@noop {} {\bibfield
   {journal} {\bibinfo  {journal} {Engineering Applications of Computational
  Fluid Mechanics}\ }\textbf {\bibinfo {volume} {17}},\ \bibinfo {pages}
  {2237611} (\bibinfo {year} {2023})}\BibitemShut {NoStop}%
\bibitem [{\citenamefont {Brunton}, \citenamefont {Noack},\ and\ \citenamefont
  {Koumoutsakos}(2020)}]{brunton2020machine}%
  \BibitemOpen
  \bibfield  {author} {\bibinfo {author} {\bibfnamefont {S.~L.}\ \bibnamefont
  {Brunton}}, \bibinfo {author} {\bibfnamefont {B.~R.}\ \bibnamefont {Noack}},\
  and\ \bibinfo {author} {\bibfnamefont {P.}~\bibnamefont {Koumoutsakos}},\
  }\bibfield  {title} {\enquote {\bibinfo {title} {Machine learning for fluid
  mechanics},}\ }\href@noop {} {\bibfield  {journal} {\bibinfo  {journal}
  {Annual review of fluid mechanics}\ }\textbf {\bibinfo {volume} {52}},\
  \bibinfo {pages} {477--508} (\bibinfo {year} {2020})}\BibitemShut {NoStop}%
\bibitem [{\citenamefont {Li}\ \emph {et~al.}(2021{\natexlab{a}})\citenamefont
  {Li}, \citenamefont {Kovachki}, \citenamefont {Azizzadenesheli},
  \citenamefont {Liu}, \citenamefont {Bhattacharya}, \citenamefont {Stuart},\
  and\ \citenamefont {Anandkumar}}]{li2020fourier}%
  \BibitemOpen
  \bibfield  {author} {\bibinfo {author} {\bibfnamefont {Z.}~\bibnamefont
  {Li}}, \bibinfo {author} {\bibfnamefont {N.}~\bibnamefont {Kovachki}},
  \bibinfo {author} {\bibfnamefont {K.}~\bibnamefont {Azizzadenesheli}},
  \bibinfo {author} {\bibfnamefont {B.}~\bibnamefont {Liu}}, \bibinfo {author}
  {\bibfnamefont {K.}~\bibnamefont {Bhattacharya}}, \bibinfo {author}
  {\bibfnamefont {A.}~\bibnamefont {Stuart}},\ and\ \bibinfo {author}
  {\bibfnamefont {A.}~\bibnamefont {Anandkumar}},\ }\bibfield  {title}
  {\enquote {\bibinfo {title} {{Fourier Neural Operator for Parametric Partial
  Differential Equations}},}\ }in\ \href@noop {} {\emph {\bibinfo {booktitle}
  {International Conference on Learning Representations}}}\ (\bibinfo {year}
  {2021})\BibitemShut {NoStop}%
\bibitem [{\citenamefont {Xue}\ \emph {et~al.}(2025{\natexlab{a}})\citenamefont
  {Xue}, \citenamefont {ten Eikelder}, \citenamefont {Yang}, \citenamefont
  {Li}, \citenamefont {He}, \citenamefont {Wang},\ and\ \citenamefont
  {Coveney}}]{xue2025equivariant}%
  \BibitemOpen
  \bibfield  {author} {\bibinfo {author} {\bibfnamefont {X.}~\bibnamefont
  {Xue}}, \bibinfo {author} {\bibfnamefont {M.~F.~P.}\ \bibnamefont {ten
  Eikelder}}, \bibinfo {author} {\bibfnamefont {T.}~\bibnamefont {Yang}},
  \bibinfo {author} {\bibfnamefont {Y.}~\bibnamefont {Li}}, \bibinfo {author}
  {\bibfnamefont {K.}~\bibnamefont {He}}, \bibinfo {author} {\bibfnamefont
  {S.}~\bibnamefont {Wang}},\ and\ \bibinfo {author} {\bibfnamefont {P.~V.}\
  \bibnamefont {Coveney}},\ }\bibfield  {title} {\enquote {\bibinfo {title}
  {{Equivariant U-Shaped Neural Operators for the Cahn-Hilliard Phase-Field
  Model}},}\ }\href@noop {} {\bibfield  {journal} {\bibinfo  {journal} {arXiv
  preprint arXiv:2509.01293}\ } (\bibinfo {year}
  {2025}{\natexlab{a}})}\BibitemShut {NoStop}%
\bibitem [{\citenamefont {Karniadakis}\ \emph {et~al.}(2021)\citenamefont
  {Karniadakis}, \citenamefont {Kevrekidis}, \citenamefont {Lu}, \citenamefont
  {Perdikaris}, \citenamefont {Wang},\ and\ \citenamefont
  {Yang}}]{karniadakis2021physics}%
  \BibitemOpen
  \bibfield  {author} {\bibinfo {author} {\bibfnamefont {G.~E.}\ \bibnamefont
  {Karniadakis}}, \bibinfo {author} {\bibfnamefont {I.~G.}\ \bibnamefont
  {Kevrekidis}}, \bibinfo {author} {\bibfnamefont {L.}~\bibnamefont {Lu}},
  \bibinfo {author} {\bibfnamefont {P.}~\bibnamefont {Perdikaris}}, \bibinfo
  {author} {\bibfnamefont {S.}~\bibnamefont {Wang}},\ and\ \bibinfo {author}
  {\bibfnamefont {L.}~\bibnamefont {Yang}},\ }\bibfield  {title} {\enquote
  {\bibinfo {title} {Physics-informed machine learning},}\ }\href@noop {}
  {\bibfield  {journal} {\bibinfo  {journal} {Nature Reviews Physics}\ }\textbf
  {\bibinfo {volume} {3}},\ \bibinfo {pages} {422--440} (\bibinfo {year}
  {2021})}\BibitemShut {NoStop}%
\bibitem [{\citenamefont {Cheng}\ \emph {et~al.}(2025)\citenamefont {Cheng},
  \citenamefont {Bocquet}, \citenamefont {Ding}, \citenamefont {Finn},
  \citenamefont {Fu}, \citenamefont {Fu}, \citenamefont {Guo}, \citenamefont
  {Johnson}, \citenamefont {Li}, \citenamefont {Liu} \emph
  {et~al.}}]{cheng2025machine}%
  \BibitemOpen
  \bibfield  {author} {\bibinfo {author} {\bibfnamefont {S.}~\bibnamefont
  {Cheng}}, \bibinfo {author} {\bibfnamefont {M.}~\bibnamefont {Bocquet}},
  \bibinfo {author} {\bibfnamefont {W.}~\bibnamefont {Ding}}, \bibinfo {author}
  {\bibfnamefont {T.~S.}\ \bibnamefont {Finn}}, \bibinfo {author}
  {\bibfnamefont {R.}~\bibnamefont {Fu}}, \bibinfo {author} {\bibfnamefont
  {J.}~\bibnamefont {Fu}}, \bibinfo {author} {\bibfnamefont {Y.}~\bibnamefont
  {Guo}}, \bibinfo {author} {\bibfnamefont {E.}~\bibnamefont {Johnson}},
  \bibinfo {author} {\bibfnamefont {S.}~\bibnamefont {Li}}, \bibinfo {author}
  {\bibfnamefont {C.}~\bibnamefont {Liu}}, \emph {et~al.},\ }\bibfield  {title}
  {\enquote {\bibinfo {title} {Machine learning for modelling unstructured grid
  data in computational physics: a review},}\ }\href@noop {} {\bibfield
  {journal} {\bibinfo  {journal} {Information Fusion}\ ,\ \bibinfo {pages}
  {103255}} (\bibinfo {year} {2025})}\BibitemShut {NoStop}%
\bibitem [{\citenamefont {Satorras}, \citenamefont {Hoogeboom},\ and\
  \citenamefont {Welling}(2021)}]{satorras2021n}%
  \BibitemOpen
  \bibfield  {author} {\bibinfo {author} {\bibfnamefont {V.~G.}\ \bibnamefont
  {Satorras}}, \bibinfo {author} {\bibfnamefont {E.}~\bibnamefont
  {Hoogeboom}},\ and\ \bibinfo {author} {\bibfnamefont {M.}~\bibnamefont
  {Welling}},\ }\bibfield  {title} {\enquote {\bibinfo {title} {{E (n)
  equivariant graph neural networks}},}\ }in\ \href@noop {} {\emph {\bibinfo
  {booktitle} {International conference on machine learning}}}\ (\bibinfo
  {organization} {PMLR},\ \bibinfo {year} {2021})\ pp.\ \bibinfo {pages}
  {9323--9332}\BibitemShut {NoStop}%
\bibitem [{\citenamefont {Hoekstra}\ and\ \citenamefont
  {Edeling}(2026)}]{hoekstra2026reduced}%
  \BibitemOpen
  \bibfield  {author} {\bibinfo {author} {\bibfnamefont {R.}~\bibnamefont
  {Hoekstra}}\ and\ \bibinfo {author} {\bibfnamefont {W.}~\bibnamefont
  {Edeling}},\ }\bibfield  {title} {\enquote {\bibinfo {title} {Reduced subgrid
  scale terms in three-dimensional turbulence},}\ }\href@noop {} {\bibfield
  {journal} {\bibinfo  {journal} {Computer Methods in Applied Mechanics and
  Engineering}\ }\textbf {\bibinfo {volume} {449}},\ \bibinfo {pages} {118506}
  (\bibinfo {year} {2026})}\BibitemShut {NoStop}%
\bibitem [{\citenamefont {van Gastelen}, \citenamefont {Edeling},\ and\
  \citenamefont {Sanderse}(2025)}]{van2025energy}%
  \BibitemOpen
  \bibfield  {author} {\bibinfo {author} {\bibfnamefont {T.}~\bibnamefont {van
  Gastelen}}, \bibinfo {author} {\bibfnamefont {W.}~\bibnamefont {Edeling}},\
  and\ \bibinfo {author} {\bibfnamefont {B.}~\bibnamefont {Sanderse}},\
  }\bibfield  {title} {\enquote {\bibinfo {title} {{Energy-Conserving Neural
  Network Closure Model for Long-Time Accurate and Stable LES}},}\ }\href@noop
  {} {\bibfield  {journal} {\bibinfo  {journal} {arXiv preprint
  arXiv:2504.05868}\ } (\bibinfo {year} {2025})}\BibitemShut {NoStop}%
\bibitem [{\citenamefont {Beck}, \citenamefont {Flad},\ and\ \citenamefont
  {Munz}(2019)}]{beck2019deep}%
  \BibitemOpen
  \bibfield  {author} {\bibinfo {author} {\bibfnamefont {A.}~\bibnamefont
  {Beck}}, \bibinfo {author} {\bibfnamefont {D.}~\bibnamefont {Flad}},\ and\
  \bibinfo {author} {\bibfnamefont {C.-D.}\ \bibnamefont {Munz}},\ }\bibfield
  {title} {\enquote {\bibinfo {title} {Deep neural networks for data-driven les
  closure models},}\ }\href@noop {} {\bibfield  {journal} {\bibinfo  {journal}
  {Journal of Computational Physics}\ }\textbf {\bibinfo {volume} {398}},\
  \bibinfo {pages} {108910} (\bibinfo {year} {2019})}\BibitemShut {NoStop}%
\bibitem [{\citenamefont {Duraisamy}, \citenamefont {Iaccarino},\ and\
  \citenamefont {Xiao}(2019)}]{duraisamy2019turbulence}%
  \BibitemOpen
  \bibfield  {author} {\bibinfo {author} {\bibfnamefont {K.}~\bibnamefont
  {Duraisamy}}, \bibinfo {author} {\bibfnamefont {G.}~\bibnamefont
  {Iaccarino}},\ and\ \bibinfo {author} {\bibfnamefont {H.}~\bibnamefont
  {Xiao}},\ }\bibfield  {title} {\enquote {\bibinfo {title} {Turbulence
  modeling in the age of data},}\ }\href@noop {} {\bibfield  {journal}
  {\bibinfo  {journal} {Annual Review of Fluid Mechanics}\ }\textbf {\bibinfo
  {volume} {51}},\ \bibinfo {pages} {357--377} (\bibinfo {year}
  {2019})}\BibitemShut {NoStop}%
\bibitem [{\citenamefont {Sarghini}, \citenamefont {De~Felice},\ and\
  \citenamefont {Santini}(2003)}]{sarghini2003neural}%
  \BibitemOpen
  \bibfield  {author} {\bibinfo {author} {\bibfnamefont {F.}~\bibnamefont
  {Sarghini}}, \bibinfo {author} {\bibfnamefont {G.}~\bibnamefont
  {De~Felice}},\ and\ \bibinfo {author} {\bibfnamefont {S.}~\bibnamefont
  {Santini}},\ }\bibfield  {title} {\enquote {\bibinfo {title} {Neural networks
  based subgrid scale modeling in large eddy simulations},}\ }\href@noop {}
  {\bibfield  {journal} {\bibinfo  {journal} {Computers \& fluids}\ }\textbf
  {\bibinfo {volume} {32}},\ \bibinfo {pages} {97--108} (\bibinfo {year}
  {2003})}\BibitemShut {NoStop}%
\bibitem [{\citenamefont {Gamahara}\ and\ \citenamefont
  {Hattori}(2017)}]{gamahara2017searching}%
  \BibitemOpen
  \bibfield  {author} {\bibinfo {author} {\bibfnamefont {M.}~\bibnamefont
  {Gamahara}}\ and\ \bibinfo {author} {\bibfnamefont {Y.}~\bibnamefont
  {Hattori}},\ }\bibfield  {title} {\enquote {\bibinfo {title} {Searching for
  turbulence models by artificial neural network},}\ }\href@noop {} {\bibfield
  {journal} {\bibinfo  {journal} {Physical Review Fluids}\ }\textbf {\bibinfo
  {volume} {2}},\ \bibinfo {pages} {054604} (\bibinfo {year}
  {2017})}\BibitemShut {NoStop}%
\bibitem [{\citenamefont {Xie}\ \emph {et~al.}(2019)\citenamefont {Xie},
  \citenamefont {Wang}, \citenamefont {Li}, \citenamefont {Wan},\ and\
  \citenamefont {Chen}}]{xie2019artificial}%
  \BibitemOpen
  \bibfield  {author} {\bibinfo {author} {\bibfnamefont {C.}~\bibnamefont
  {Xie}}, \bibinfo {author} {\bibfnamefont {J.}~\bibnamefont {Wang}}, \bibinfo
  {author} {\bibfnamefont {H.}~\bibnamefont {Li}}, \bibinfo {author}
  {\bibfnamefont {M.}~\bibnamefont {Wan}},\ and\ \bibinfo {author}
  {\bibfnamefont {S.}~\bibnamefont {Chen}},\ }\bibfield  {title} {\enquote
  {\bibinfo {title} {Artificial neural network mixed model for large eddy
  simulation of compressible isotropic turbulence},}\ }\href@noop {} {\bibfield
   {journal} {\bibinfo  {journal} {Physics of Fluids}\ }\textbf {\bibinfo
  {volume} {31}} (\bibinfo {year} {2019})}\BibitemShut {NoStop}%
\bibitem [{\citenamefont {Mohan}\ \emph {et~al.}(2023)\citenamefont {Mohan},
  \citenamefont {Lubbers}, \citenamefont {Chertkov},\ and\ \citenamefont
  {Livescu}}]{mohan2023embedding}%
  \BibitemOpen
  \bibfield  {author} {\bibinfo {author} {\bibfnamefont {A.~T.}\ \bibnamefont
  {Mohan}}, \bibinfo {author} {\bibfnamefont {N.}~\bibnamefont {Lubbers}},
  \bibinfo {author} {\bibfnamefont {M.}~\bibnamefont {Chertkov}},\ and\
  \bibinfo {author} {\bibfnamefont {D.}~\bibnamefont {Livescu}},\ }\bibfield
  {title} {\enquote {\bibinfo {title} {Embedding hard physical constraints in
  neural network coarse-graining of three-dimensional turbulence},}\
  }\href@noop {} {\bibfield  {journal} {\bibinfo  {journal} {Physical Review
  Fluids}\ }\textbf {\bibinfo {volume} {8}},\ \bibinfo {pages} {014604}
  (\bibinfo {year} {2023})}\BibitemShut {NoStop}%
\bibitem [{\citenamefont {Fischer}\ \emph {et~al.}(2025)\citenamefont
  {Fischer}, \citenamefont {Kaltenbach}, \citenamefont {Litvinov},
  \citenamefont {Succi},\ and\ \citenamefont
  {Koumoutsakos}}]{fischer2025optimal}%
  \BibitemOpen
  \bibfield  {author} {\bibinfo {author} {\bibfnamefont {P.}~\bibnamefont
  {Fischer}}, \bibinfo {author} {\bibfnamefont {S.}~\bibnamefont {Kaltenbach}},
  \bibinfo {author} {\bibfnamefont {S.}~\bibnamefont {Litvinov}}, \bibinfo
  {author} {\bibfnamefont {S.}~\bibnamefont {Succi}},\ and\ \bibinfo {author}
  {\bibfnamefont {P.}~\bibnamefont {Koumoutsakos}},\ }\bibfield  {title}
  {\enquote {\bibinfo {title} {Optimal lattice boltzmann closures through
  multi-agent reinforcement learning},}\ }\href@noop {} {\bibfield  {journal}
  {\bibinfo  {journal} {arXiv preprint arXiv:2504.14422}\ } (\bibinfo {year}
  {2025})}\BibitemShut {NoStop}%
\bibitem [{\citenamefont {Sherstinsky}(2020)}]{sherstinsky2020fundamentals}%
  \BibitemOpen
  \bibfield  {author} {\bibinfo {author} {\bibfnamefont {A.}~\bibnamefont
  {Sherstinsky}},\ }\bibfield  {title} {\enquote {\bibinfo {title}
  {Fundamentals of recurrent neural network (rnn) and long short-term memory
  (lstm) network},}\ }\href@noop {} {\bibfield  {journal} {\bibinfo  {journal}
  {Physica D: Nonlinear Phenomena}\ }\textbf {\bibinfo {volume} {404}},\
  \bibinfo {pages} {132306} (\bibinfo {year} {2020})}\BibitemShut {NoStop}%
\bibitem [{\citenamefont {Greff}\ \emph {et~al.}(2016)\citenamefont {Greff},
  \citenamefont {Srivastava}, \citenamefont {Koutn{\'\i}k}, \citenamefont
  {Steunebrink},\ and\ \citenamefont {Schmidhuber}}]{greff2016lstm}%
  \BibitemOpen
  \bibfield  {author} {\bibinfo {author} {\bibfnamefont {K.}~\bibnamefont
  {Greff}}, \bibinfo {author} {\bibfnamefont {R.~K.}\ \bibnamefont
  {Srivastava}}, \bibinfo {author} {\bibfnamefont {J.}~\bibnamefont
  {Koutn{\'\i}k}}, \bibinfo {author} {\bibfnamefont {B.~R.}\ \bibnamefont
  {Steunebrink}},\ and\ \bibinfo {author} {\bibfnamefont {J.}~\bibnamefont
  {Schmidhuber}},\ }\bibfield  {title} {\enquote {\bibinfo {title} {{LSTM: A
  search space odyssey}},}\ }\href@noop {} {\bibfield  {journal} {\bibinfo
  {journal} {IEEE transactions on neural networks and learning systems}\
  }\textbf {\bibinfo {volume} {28}},\ \bibinfo {pages} {2222--2232} (\bibinfo
  {year} {2016})}\BibitemShut {NoStop}%
\bibitem [{\citenamefont {Lu}\ \emph {et~al.}(2021)\citenamefont {Lu},
  \citenamefont {Jin}, \citenamefont {Pang}, \citenamefont {Zhang},\ and\
  \citenamefont {Karniadakis}}]{lu2021learning}%
  \BibitemOpen
  \bibfield  {author} {\bibinfo {author} {\bibfnamefont {L.}~\bibnamefont
  {Lu}}, \bibinfo {author} {\bibfnamefont {P.}~\bibnamefont {Jin}}, \bibinfo
  {author} {\bibfnamefont {G.}~\bibnamefont {Pang}}, \bibinfo {author}
  {\bibfnamefont {Z.}~\bibnamefont {Zhang}},\ and\ \bibinfo {author}
  {\bibfnamefont {G.~E.}\ \bibnamefont {Karniadakis}},\ }\bibfield  {title}
  {\enquote {\bibinfo {title} {Learning nonlinear operators via deeponet based
  on the universal approximation theorem of operators},}\ }\href@noop {}
  {\bibfield  {journal} {\bibinfo  {journal} {Nature Machine Intelligence}\
  }\textbf {\bibinfo {volume} {3}},\ \bibinfo {pages} {218--229} (\bibinfo
  {year} {2021})}\BibitemShut {NoStop}%
\bibitem [{\citenamefont {Kochkov}\ \emph {et~al.}(2021)\citenamefont
  {Kochkov}, \citenamefont {Smith}, \citenamefont {Alieva}, \citenamefont
  {Wang}, \citenamefont {Brenner},\ and\ \citenamefont
  {Hoyer}}]{kochkov2021machine}%
  \BibitemOpen
  \bibfield  {author} {\bibinfo {author} {\bibfnamefont {D.}~\bibnamefont
  {Kochkov}}, \bibinfo {author} {\bibfnamefont {J.~A.}\ \bibnamefont {Smith}},
  \bibinfo {author} {\bibfnamefont {A.}~\bibnamefont {Alieva}}, \bibinfo
  {author} {\bibfnamefont {Q.}~\bibnamefont {Wang}}, \bibinfo {author}
  {\bibfnamefont {M.~P.}\ \bibnamefont {Brenner}},\ and\ \bibinfo {author}
  {\bibfnamefont {S.}~\bibnamefont {Hoyer}},\ }\bibfield  {title} {\enquote
  {\bibinfo {title} {Machine learning--accelerated computational fluid
  dynamics},}\ }\href@noop {} {\bibfield  {journal} {\bibinfo  {journal}
  {Proceedings of the National Academy of Sciences}\ }\textbf {\bibinfo
  {volume} {118}},\ \bibinfo {pages} {e2101784118} (\bibinfo {year}
  {2021})}\BibitemShut {NoStop}%
\bibitem [{\citenamefont {Ye}, \citenamefont {Zhang},\ and\ \citenamefont
  {Wang}(2025)}]{ye2025recurrent}%
  \BibitemOpen
  \bibfield  {author} {\bibinfo {author} {\bibfnamefont {Z.}~\bibnamefont
  {Ye}}, \bibinfo {author} {\bibfnamefont {C.-S.}\ \bibnamefont {Zhang}},\ and\
  \bibinfo {author} {\bibfnamefont {W.}~\bibnamefont {Wang}},\ }\bibfield
  {title} {\enquote {\bibinfo {title} {{Recurrent Neural Operators: Stable
  Long-Term PDE Prediction}},}\ }\href@noop {} {\bibfield  {journal} {\bibinfo
  {journal} {arXiv preprint arXiv:2505.20721}\ } (\bibinfo {year}
  {2025})}\BibitemShut {NoStop}%
\bibitem [{\citenamefont {Wang}\ \emph {et~al.}(2025)\citenamefont {Wang},
  \citenamefont {Xue}, \citenamefont {Gao},\ and\ \citenamefont
  {Coveney}}]{wangxue2025quantum}%
  \BibitemOpen
  \bibfield  {author} {\bibinfo {author} {\bibfnamefont {M.}~\bibnamefont
  {Wang}}, \bibinfo {author} {\bibfnamefont {X.}~\bibnamefont {Xue}}, \bibinfo
  {author} {\bibfnamefont {M.}~\bibnamefont {Gao}},\ and\ \bibinfo {author}
  {\bibfnamefont {P.~V.}\ \bibnamefont {Coveney}},\ }\bibfield  {title}
  {\enquote {\bibinfo {title} {{Quantum-Informed Machine Learning for
  Predicting Spatiotemporal Chaos}},}\ }\href@noop {} {\bibfield  {journal}
  {\bibinfo  {journal} {arXiv preprint arXiv:2507.19861}\ } (\bibinfo {year}
  {2025})}\BibitemShut {NoStop}%
\bibitem [{\citenamefont {McCabe}\ \emph {et~al.}()\citenamefont {McCabe},
  \citenamefont {Harrington}, \citenamefont {Subramanian},\ and\ \citenamefont
  {Brown}}]{mccabetowards}%
  \BibitemOpen
  \bibfield  {author} {\bibinfo {author} {\bibfnamefont {M.}~\bibnamefont
  {McCabe}}, \bibinfo {author} {\bibfnamefont {P.}~\bibnamefont {Harrington}},
  \bibinfo {author} {\bibfnamefont {S.}~\bibnamefont {Subramanian}},\ and\
  \bibinfo {author} {\bibfnamefont {J.}~\bibnamefont {Brown}},\ }\bibfield
  {title} {\enquote {\bibinfo {title} {Towards stability of autoregressive
  neural operators},}\ }\href@noop {} {\bibinfo  {journal} {Transactions on
  Machine Learning Research}\ }\BibitemShut {NoStop}%
\bibitem [{\citenamefont {Song}\ \emph {et~al.}(2020)\citenamefont {Song},
  \citenamefont {Sohl-Dickstein}, \citenamefont {Kingma}, \citenamefont
  {Kumar}, \citenamefont {Ermon},\ and\ \citenamefont {Poole}}]{song2020score}%
  \BibitemOpen
\bibfield  {journal} {  }\bibfield  {author} {\bibinfo {author} {\bibfnamefont
  {Y.}~\bibnamefont {Song}}, \bibinfo {author} {\bibfnamefont {J.}~\bibnamefont
  {Sohl-Dickstein}}, \bibinfo {author} {\bibfnamefont {D.~P.}\ \bibnamefont
  {Kingma}}, \bibinfo {author} {\bibfnamefont {A.}~\bibnamefont {Kumar}},
  \bibinfo {author} {\bibfnamefont {S.}~\bibnamefont {Ermon}},\ and\ \bibinfo
  {author} {\bibfnamefont {B.}~\bibnamefont {Poole}},\ }\bibfield  {title}
  {\enquote {\bibinfo {title} {Score-based generative modeling through
  stochastic differential equations},}\ }\href@noop {} {\bibfield  {journal}
  {\bibinfo  {journal} {arXiv preprint arXiv:2011.13456}\ } (\bibinfo {year}
  {2020})}\BibitemShut {NoStop}%
\bibitem [{\citenamefont {Ho}, \citenamefont {Jain},\ and\ \citenamefont
  {Abbeel}(2020)}]{ho2020ddpm}%
  \BibitemOpen
  \bibfield  {author} {\bibinfo {author} {\bibfnamefont {J.}~\bibnamefont
  {Ho}}, \bibinfo {author} {\bibfnamefont {A.}~\bibnamefont {Jain}},\ and\
  \bibinfo {author} {\bibfnamefont {P.}~\bibnamefont {Abbeel}},\ }\bibfield
  {title} {\enquote {\bibinfo {title} {Denoising diffusion probabilistic
  models},}\ }\href@noop {} {\bibfield  {journal} {\bibinfo  {journal}
  {Advances in neural information processing systems}\ }\textbf {\bibinfo
  {volume} {33}},\ \bibinfo {pages} {6840--6851} (\bibinfo {year}
  {2020})}\BibitemShut {NoStop}%
\bibitem [{\citenamefont {Mokady}\ \emph {et~al.}(2023)\citenamefont {Mokady},
  \citenamefont {Hertz}, \citenamefont {Aberman}, \citenamefont {Pritch},\ and\
  \citenamefont {Cohen-Or}}]{mokady2022nulltextinversioneditingreal}%
  \BibitemOpen
  \bibfield  {author} {\bibinfo {author} {\bibfnamefont {R.}~\bibnamefont
  {Mokady}}, \bibinfo {author} {\bibfnamefont {A.}~\bibnamefont {Hertz}},
  \bibinfo {author} {\bibfnamefont {K.}~\bibnamefont {Aberman}}, \bibinfo
  {author} {\bibfnamefont {Y.}~\bibnamefont {Pritch}},\ and\ \bibinfo {author}
  {\bibfnamefont {D.}~\bibnamefont {Cohen-Or}},\ }\bibfield  {title} {\enquote
  {\bibinfo {title} {Null-text inversion for editing real images using guided
  diffusion models},}\ }in\ \href@noop {} {\emph {\bibinfo {booktitle}
  {Proceedings of the IEEE/CVF conference on computer vision and pattern
  recognition}}}\ (\bibinfo {year} {2023})\ pp.\ \bibinfo {pages}
  {6038--6047}\BibitemShut {NoStop}%
\bibitem [{\citenamefont {Esser}, \citenamefont {Rombach},\ and\ \citenamefont
  {Ommer}(2021)}]{esser2021tamingtransformer}%
  \BibitemOpen
  \bibfield  {author} {\bibinfo {author} {\bibfnamefont {P.}~\bibnamefont
  {Esser}}, \bibinfo {author} {\bibfnamefont {R.}~\bibnamefont {Rombach}},\
  and\ \bibinfo {author} {\bibfnamefont {B.}~\bibnamefont {Ommer}},\ }\bibfield
   {title} {\enquote {\bibinfo {title} {Taming transformers for high-resolution
  image synthesis},}\ }in\ \href@noop {} {\emph {\bibinfo {booktitle}
  {Proceedings of the IEEE/CVF conference on computer vision and pattern
  recognition}}}\ (\bibinfo {year} {2021})\ pp.\ \bibinfo {pages}
  {12873--12883}\BibitemShut {NoStop}%
\bibitem [{\citenamefont {Lam}\ \emph {et~al.}(2023)\citenamefont {Lam},
  \citenamefont {Sanchez-Gonzalez}, \citenamefont {Willson}, \citenamefont
  {Wirnsberger}, \citenamefont {Fortunato}, \citenamefont {Alet}, \citenamefont
  {Ravuri}, \citenamefont {Ewalds}, \citenamefont {Eaton-Rosen}, \citenamefont
  {Hu} \emph {et~al.}}]{lam2023learning}%
  \BibitemOpen
  \bibfield  {author} {\bibinfo {author} {\bibfnamefont {R.}~\bibnamefont
  {Lam}}, \bibinfo {author} {\bibfnamefont {A.}~\bibnamefont
  {Sanchez-Gonzalez}}, \bibinfo {author} {\bibfnamefont {M.}~\bibnamefont
  {Willson}}, \bibinfo {author} {\bibfnamefont {P.}~\bibnamefont
  {Wirnsberger}}, \bibinfo {author} {\bibfnamefont {M.}~\bibnamefont
  {Fortunato}}, \bibinfo {author} {\bibfnamefont {F.}~\bibnamefont {Alet}},
  \bibinfo {author} {\bibfnamefont {S.}~\bibnamefont {Ravuri}}, \bibinfo
  {author} {\bibfnamefont {T.}~\bibnamefont {Ewalds}}, \bibinfo {author}
  {\bibfnamefont {Z.}~\bibnamefont {Eaton-Rosen}}, \bibinfo {author}
  {\bibfnamefont {W.}~\bibnamefont {Hu}}, \emph {et~al.},\ }\bibfield  {title}
  {\enquote {\bibinfo {title} {Learning skillful medium-range global weather
  forecasting},}\ }\href@noop {} {\bibfield  {journal} {\bibinfo  {journal}
  {Science}\ }\textbf {\bibinfo {volume} {382}},\ \bibinfo {pages} {1416--1421}
  (\bibinfo {year} {2023})}\BibitemShut {NoStop}%
\bibitem [{\citenamefont {R{\"u}hling~Cachay}\ \emph
  {et~al.}(2023)\citenamefont {R{\"u}hling~Cachay}, \citenamefont {Zhao},
  \citenamefont {Joren},\ and\ \citenamefont {Yu}}]{ruhling2023dyffusion}%
  \BibitemOpen
  \bibfield  {author} {\bibinfo {author} {\bibfnamefont {S.}~\bibnamefont
  {R{\"u}hling~Cachay}}, \bibinfo {author} {\bibfnamefont {B.}~\bibnamefont
  {Zhao}}, \bibinfo {author} {\bibfnamefont {H.}~\bibnamefont {Joren}},\ and\
  \bibinfo {author} {\bibfnamefont {R.}~\bibnamefont {Yu}},\ }\bibfield
  {title} {\enquote {\bibinfo {title} {Dyffusion: A dynamics-informed diffusion
  model for spatiotemporal forecasting},}\ }\href@noop {} {\bibfield  {journal}
  {\bibinfo  {journal} {Advances in neural information processing systems}\
  }\textbf {\bibinfo {volume} {36}},\ \bibinfo {pages} {45259--45287} (\bibinfo
  {year} {2023})}\BibitemShut {NoStop}%
\bibitem [{\citenamefont {Rombach}\ \emph {et~al.}(2022)\citenamefont
  {Rombach}, \citenamefont {Blattmann}, \citenamefont {Lorenz}, \citenamefont
  {Esser},\ and\ \citenamefont {Ommer}}]{rombach2022high}%
  \BibitemOpen
  \bibfield  {author} {\bibinfo {author} {\bibfnamefont {R.}~\bibnamefont
  {Rombach}}, \bibinfo {author} {\bibfnamefont {A.}~\bibnamefont {Blattmann}},
  \bibinfo {author} {\bibfnamefont {D.}~\bibnamefont {Lorenz}}, \bibinfo
  {author} {\bibfnamefont {P.}~\bibnamefont {Esser}},\ and\ \bibinfo {author}
  {\bibfnamefont {B.}~\bibnamefont {Ommer}},\ }\bibfield  {title} {\enquote
  {\bibinfo {title} {High-resolution image synthesis with latent diffusion
  models},}\ }in\ \href@noop {} {\emph {\bibinfo {booktitle} {Proceedings of
  the IEEE/CVF conference on computer vision and pattern recognition}}}\
  (\bibinfo {year} {2022})\ pp.\ \bibinfo {pages} {10684--10695}\BibitemShut
  {NoStop}%
\bibitem [{\citenamefont {Du}\ \emph {et~al.}(2024)\citenamefont {Du},
  \citenamefont {Parikh}, \citenamefont {Fan}, \citenamefont {Liu},\ and\
  \citenamefont {Wang}}]{du2024conditional}%
  \BibitemOpen
  \bibfield  {author} {\bibinfo {author} {\bibfnamefont {P.}~\bibnamefont
  {Du}}, \bibinfo {author} {\bibfnamefont {M.~H.}\ \bibnamefont {Parikh}},
  \bibinfo {author} {\bibfnamefont {X.}~\bibnamefont {Fan}}, \bibinfo {author}
  {\bibfnamefont {X.-Y.}\ \bibnamefont {Liu}},\ and\ \bibinfo {author}
  {\bibfnamefont {J.-X.}\ \bibnamefont {Wang}},\ }\bibfield  {title} {\enquote
  {\bibinfo {title} {Conditional neural field latent diffusion model for
  generating spatiotemporal turbulence},}\ }\href@noop {} {\bibfield  {journal}
  {\bibinfo  {journal} {Nature Communications}\ }\textbf {\bibinfo {volume}
  {15}},\ \bibinfo {pages} {10416} (\bibinfo {year} {2024})}\BibitemShut
  {NoStop}%
\bibitem [{\citenamefont {Zacharoudiou}, \citenamefont {McCullough},\ and\
  \citenamefont {Coveney}(2023)}]{zacharoudiou2023development}%
  \BibitemOpen
  \bibfield  {author} {\bibinfo {author} {\bibfnamefont {I.}~\bibnamefont
  {Zacharoudiou}}, \bibinfo {author} {\bibfnamefont {J.~W.~S.}\ \bibnamefont
  {McCullough}},\ and\ \bibinfo {author} {\bibfnamefont {P.~V.}\ \bibnamefont
  {Coveney}},\ }\bibfield  {title} {\enquote {\bibinfo {title} {{Development
  and performance of a HemeLB GPU code for human-scale blood flow
  simulation}},}\ }\href@noop {} {\bibfield  {journal} {\bibinfo  {journal}
  {Computer Physics Communications}\ }\textbf {\bibinfo {volume} {282}},\
  \bibinfo {pages} {108548} (\bibinfo {year} {2023})}\BibitemShut {NoStop}%
\bibitem [{\citenamefont {N{\o}rgaard}\ \emph {et~al.}(2014)\citenamefont
  {N{\o}rgaard}, \citenamefont {Leipsic}, \citenamefont {Gaur}, \citenamefont
  {Seneviratne}, \citenamefont {Ko}, \citenamefont {Ito}, \citenamefont
  {Jensen}, \citenamefont {Mauri}, \citenamefont {De~Bruyne}, \citenamefont
  {Bezerra} \emph {et~al.}}]{norgaard2014diagnostic}%
  \BibitemOpen
  \bibfield  {author} {\bibinfo {author} {\bibfnamefont {B.~L.}\ \bibnamefont
  {N{\o}rgaard}}, \bibinfo {author} {\bibfnamefont {J.}~\bibnamefont
  {Leipsic}}, \bibinfo {author} {\bibfnamefont {S.}~\bibnamefont {Gaur}},
  \bibinfo {author} {\bibfnamefont {S.}~\bibnamefont {Seneviratne}}, \bibinfo
  {author} {\bibfnamefont {B.~S.}\ \bibnamefont {Ko}}, \bibinfo {author}
  {\bibfnamefont {H.}~\bibnamefont {Ito}}, \bibinfo {author} {\bibfnamefont
  {J.~M.}\ \bibnamefont {Jensen}}, \bibinfo {author} {\bibfnamefont
  {L.}~\bibnamefont {Mauri}}, \bibinfo {author} {\bibfnamefont
  {B.}~\bibnamefont {De~Bruyne}}, \bibinfo {author} {\bibfnamefont
  {H.}~\bibnamefont {Bezerra}}, \emph {et~al.},\ }\bibfield  {title} {\enquote
  {\bibinfo {title} {{Diagnostic performance of noninvasive fractional flow
  reserve derived from coronary computed tomography angiography in suspected
  coronary artery disease: the NXT trial (Analysis of Coronary Blood Flow Using
  CT Angiography: Next Steps)}},}\ }\href@noop {} {\bibfield  {journal}
  {\bibinfo  {journal} {Journal of the American College of Cardiology}\
  }\textbf {\bibinfo {volume} {63}},\ \bibinfo {pages} {1145--1155} (\bibinfo
  {year} {2014})}\BibitemShut {NoStop}%
\bibitem [{\citenamefont {Tanade}\ \emph {et~al.}(2022)\citenamefont {Tanade},
  \citenamefont {Chen}, \citenamefont {Leopold},\ and\ \citenamefont
  {Randles}}]{tanade2022analysis}%
  \BibitemOpen
  \bibfield  {author} {\bibinfo {author} {\bibfnamefont {C.}~\bibnamefont
  {Tanade}}, \bibinfo {author} {\bibfnamefont {S.~J.}\ \bibnamefont {Chen}},
  \bibinfo {author} {\bibfnamefont {J.~A.}\ \bibnamefont {Leopold}},\ and\
  \bibinfo {author} {\bibfnamefont {A.}~\bibnamefont {Randles}},\ }\bibfield
  {title} {\enquote {\bibinfo {title} {Analysis identifying minimal governing
  parameters for clinically accurate in silico fractional flow reserve},}\
  }\href@noop {} {\bibfield  {journal} {\bibinfo  {journal} {Frontiers in
  Medical Technology}\ }\textbf {\bibinfo {volume} {4}},\ \bibinfo {pages}
  {1034801} (\bibinfo {year} {2022})}\BibitemShut {NoStop}%
\bibitem [{\citenamefont {Lund}, \citenamefont {Wu},\ and\ \citenamefont
  {Squires}(1998)}]{lund1998generation}%
  \BibitemOpen
  \bibfield  {author} {\bibinfo {author} {\bibfnamefont {T.~S.}\ \bibnamefont
  {Lund}}, \bibinfo {author} {\bibfnamefont {X.}~\bibnamefont {Wu}},\ and\
  \bibinfo {author} {\bibfnamefont {K.~D.}\ \bibnamefont {Squires}},\
  }\bibfield  {title} {\enquote {\bibinfo {title} {Generation of turbulent
  inflow data for spatially-developing boundary layer simulations},}\
  }\href@noop {} {\bibfield  {journal} {\bibinfo  {journal} {Journal of
  computational physics}\ }\textbf {\bibinfo {volume} {140}},\ \bibinfo {pages}
  {233--258} (\bibinfo {year} {1998})}\BibitemShut {NoStop}%
\bibitem [{\citenamefont {Wu}(2017)}]{wu2017inflow}%
  \BibitemOpen
  \bibfield  {author} {\bibinfo {author} {\bibfnamefont {X.}~\bibnamefont
  {Wu}},\ }\bibfield  {title} {\enquote {\bibinfo {title} {Inflow turbulence
  generation methods},}\ }\href@noop {} {\bibfield  {journal} {\bibinfo
  {journal} {Annual Review of Fluid Mechanics}\ }\textbf {\bibinfo {volume}
  {49}},\ \bibinfo {pages} {23--49} (\bibinfo {year} {2017})}\BibitemShut
  {NoStop}%
\bibitem [{\citenamefont {Coveney}\ \emph {et~al.}(2025)\citenamefont
  {Coveney}, \citenamefont {Highfield}, \citenamefont {Stahlberg},\ and\
  \citenamefont {V{\'a}zquez}}]{coveney2025digital}%
  \BibitemOpen
  \bibfield  {author} {\bibinfo {author} {\bibfnamefont {P.}~\bibnamefont
  {Coveney}}, \bibinfo {author} {\bibfnamefont {R.}~\bibnamefont {Highfield}},
  \bibinfo {author} {\bibfnamefont {E.}~\bibnamefont {Stahlberg}},\ and\
  \bibinfo {author} {\bibfnamefont {M.}~\bibnamefont {V{\'a}zquez}},\
  }\bibfield  {title} {\enquote {\bibinfo {title} {{Digital twins and Big AI:
  the future of truly individualised healthcare}},}\ }\href@noop {} {\bibfield
  {journal} {\bibinfo  {journal} {NPJ Digital Medicine}\ }\textbf {\bibinfo
  {volume} {8}},\ \bibinfo {pages} {494} (\bibinfo {year} {2025})}\BibitemShut
  {NoStop}%
\bibitem [{\citenamefont {Schl{\"u}ter}, \citenamefont {Pitsch},\ and\
  \citenamefont {Moin}(2004)}]{schluter2004large}%
  \BibitemOpen
  \bibfield  {author} {\bibinfo {author} {\bibfnamefont {J.}~\bibnamefont
  {Schl{\"u}ter}}, \bibinfo {author} {\bibfnamefont {H.}~\bibnamefont
  {Pitsch}},\ and\ \bibinfo {author} {\bibfnamefont {P.}~\bibnamefont {Moin}},\
  }\bibfield  {title} {\enquote {\bibinfo {title} {Large-eddy simulation inflow
  conditions for coupling with reynolds-averaged flow solvers},}\ }\href@noop
  {} {\bibfield  {journal} {\bibinfo  {journal} {AIAA journal}\ }\textbf
  {\bibinfo {volume} {42}},\ \bibinfo {pages} {478--484} (\bibinfo {year}
  {2004})}\BibitemShut {NoStop}%
\bibitem [{\citenamefont {Poletto}\ \emph {et~al.}(2011)\citenamefont
  {Poletto}, \citenamefont {Revell}, \citenamefont {Craft},\ and\ \citenamefont
  {Jarrin}}]{poletto2011divergence}%
  \BibitemOpen
  \bibfield  {author} {\bibinfo {author} {\bibfnamefont {R.}~\bibnamefont
  {Poletto}}, \bibinfo {author} {\bibfnamefont {A.}~\bibnamefont {Revell}},
  \bibinfo {author} {\bibfnamefont {T.~J.}\ \bibnamefont {Craft}},\ and\
  \bibinfo {author} {\bibfnamefont {N.}~\bibnamefont {Jarrin}},\ }\bibfield
  {title} {\enquote {\bibinfo {title} {Divergence free synthetic eddy method
  for embedded les inflow boundary conditions},}\ }in\ \href@noop {} {\emph
  {\bibinfo {booktitle} {Seventh international symposium on turbulence and
  shear flow phenomena}}}\ (\bibinfo {organization} {Begel House Inc.},\
  \bibinfo {year} {2011})\BibitemShut {NoStop}%
\bibitem [{\citenamefont {Xue}\ \emph {et~al.}(2025{\natexlab{b}})\citenamefont
  {Xue}, \citenamefont {Athawale}, \citenamefont {McCullough}, \citenamefont
  {Lo}, \citenamefont {Zacharoudiou}, \citenamefont {Joo}, \citenamefont
  {Georgiadou},\ and\ \citenamefont {Coveney}}]{xue2025uncertainty}%
  \BibitemOpen
  \bibfield  {author} {\bibinfo {author} {\bibfnamefont {X.}~\bibnamefont
  {Xue}}, \bibinfo {author} {\bibfnamefont {T.~M.}\ \bibnamefont {Athawale}},
  \bibinfo {author} {\bibfnamefont {J.~W.}\ \bibnamefont {McCullough}},
  \bibinfo {author} {\bibfnamefont {S.~C.}\ \bibnamefont {Lo}}, \bibinfo
  {author} {\bibfnamefont {I.}~\bibnamefont {Zacharoudiou}}, \bibinfo {author}
  {\bibfnamefont {B.}~\bibnamefont {Joo}}, \bibinfo {author} {\bibfnamefont
  {A.}~\bibnamefont {Georgiadou}},\ and\ \bibinfo {author} {\bibfnamefont
  {P.~V.}\ \bibnamefont {Coveney}},\ }\bibfield  {title} {\enquote {\bibinfo
  {title} {An uncertainty visualization framework for large-scale
  cardiovascular flow simulations: A case study on aortic stenosis},}\
  }\href@noop {} {\bibfield  {journal} {\bibinfo  {journal} {arXiv preprint
  arXiv:2508.15420}\ } (\bibinfo {year} {2025}{\natexlab{b}})}\BibitemShut
  {NoStop}%
\bibitem [{\citenamefont {Lo}\ \emph {et~al.}(2025)\citenamefont {Lo},
  \citenamefont {Zingaro}, \citenamefont {McCullough}, \citenamefont {Xue},
  \citenamefont {Gonzalez-Martin}, \citenamefont {Joo}, \citenamefont
  {V{\'a}zquez},\ and\ \citenamefont {Coveney}}]{lo2025multi}%
  \BibitemOpen
  \bibfield  {author} {\bibinfo {author} {\bibfnamefont {S.~C.}\ \bibnamefont
  {Lo}}, \bibinfo {author} {\bibfnamefont {A.}~\bibnamefont {Zingaro}},
  \bibinfo {author} {\bibfnamefont {J.~W.}\ \bibnamefont {McCullough}},
  \bibinfo {author} {\bibfnamefont {X.}~\bibnamefont {Xue}}, \bibinfo {author}
  {\bibfnamefont {P.}~\bibnamefont {Gonzalez-Martin}}, \bibinfo {author}
  {\bibfnamefont {B.}~\bibnamefont {Joo}}, \bibinfo {author} {\bibfnamefont
  {M.}~\bibnamefont {V{\'a}zquez}},\ and\ \bibinfo {author} {\bibfnamefont
  {P.~V.}\ \bibnamefont {Coveney}},\ }\bibfield  {title} {\enquote {\bibinfo
  {title} {A multi-component, multi-physics computational model for solving
  coupled cardiac electromechanics and vascular haemodynamics},}\ }\href@noop
  {} {\bibfield  {journal} {\bibinfo  {journal} {Computer Methods in Applied
  Mechanics and Engineering}\ }\textbf {\bibinfo {volume} {446}},\ \bibinfo
  {pages} {118185} (\bibinfo {year} {2025})}\BibitemShut {NoStop}%
\bibitem [{\citenamefont {Xue}\ \emph {et~al.}(2024)\citenamefont {Xue},
  \citenamefont {McCullough}, \citenamefont {Lo}, \citenamefont {Zacharoudiou},
  \citenamefont {Jo{\'o}},\ and\ \citenamefont {Coveney}}]{xue2024lattice}%
  \BibitemOpen
  \bibfield  {author} {\bibinfo {author} {\bibfnamefont {X.}~\bibnamefont
  {Xue}}, \bibinfo {author} {\bibfnamefont {J.~W.}\ \bibnamefont {McCullough}},
  \bibinfo {author} {\bibfnamefont {S.~C.}\ \bibnamefont {Lo}}, \bibinfo
  {author} {\bibfnamefont {I.}~\bibnamefont {Zacharoudiou}}, \bibinfo {author}
  {\bibfnamefont {B.}~\bibnamefont {Jo{\'o}}},\ and\ \bibinfo {author}
  {\bibfnamefont {P.~V.}\ \bibnamefont {Coveney}},\ }\bibfield  {title}
  {\enquote {\bibinfo {title} {{The lattice Boltzmann based large eddy
  simulations for the stenosis of the aorta}},}\ }in\ \href@noop {} {\emph
  {\bibinfo {booktitle} {International Conference on Computational Science}}}\
  (\bibinfo {organization} {Springer Nature Switzerland},\ \bibinfo {year}
  {2024})\ pp.\ \bibinfo {pages} {408--420}\BibitemShut {NoStop}%
\bibitem [{\citenamefont {Mezi{\'c}}(2021)}]{mezic2021koopman}%
  \BibitemOpen
  \bibfield  {author} {\bibinfo {author} {\bibfnamefont {I.}~\bibnamefont
  {Mezi{\'c}}},\ }\bibfield  {title} {\enquote {\bibinfo {title} {Koopman
  operator, geometry, and learning of dynamical systems},}\ }\href@noop {}
  {\bibfield  {journal} {\bibinfo  {journal} {Not. Am. Math. Soc}\ }\textbf
  {\bibinfo {volume} {68}},\ \bibinfo {pages} {1087--1105} (\bibinfo {year}
  {2021})}\BibitemShut {NoStop}%
\bibitem [{\citenamefont {Brunton}\ \emph {et~al.}(2021)\citenamefont
  {Brunton}, \citenamefont {Budi{\v{s}}i{\'c}}, \citenamefont {Kaiser},\ and\
  \citenamefont {Kutz}}]{brunton2021modern}%
  \BibitemOpen
  \bibfield  {author} {\bibinfo {author} {\bibfnamefont {S.~L.}\ \bibnamefont
  {Brunton}}, \bibinfo {author} {\bibfnamefont {M.}~\bibnamefont
  {Budi{\v{s}}i{\'c}}}, \bibinfo {author} {\bibfnamefont {E.}~\bibnamefont
  {Kaiser}},\ and\ \bibinfo {author} {\bibfnamefont {J.~N.}\ \bibnamefont
  {Kutz}},\ }\bibfield  {title} {\enquote {\bibinfo {title} {{Modern Koopman
  theory for dynamical systems}},}\ }\href@noop {} {\bibfield  {journal}
  {\bibinfo  {journal} {arXiv preprint arXiv:2102.12086}\ } (\bibinfo {year}
  {2021})}\BibitemShut {NoStop}%
\bibitem [{\citenamefont {Lo}\ \emph {et~al.}(2024)\citenamefont {Lo},
  \citenamefont {Zingaro}, \citenamefont {McCullough}, \citenamefont {Xue},
  \citenamefont {V{\'a}zquez},\ and\ \citenamefont {Coveney}}]{lo2024multi}%
  \BibitemOpen
  \bibfield  {author} {\bibinfo {author} {\bibfnamefont {S.~C.}\ \bibnamefont
  {Lo}}, \bibinfo {author} {\bibfnamefont {A.}~\bibnamefont {Zingaro}},
  \bibinfo {author} {\bibfnamefont {J.~W.}\ \bibnamefont {McCullough}},
  \bibinfo {author} {\bibfnamefont {X.}~\bibnamefont {Xue}}, \bibinfo {author}
  {\bibfnamefont {M.}~\bibnamefont {V{\'a}zquez}},\ and\ \bibinfo {author}
  {\bibfnamefont {P.~V.}\ \bibnamefont {Coveney}},\ }\bibfield  {title}
  {\enquote {\bibinfo {title} {A multi-component, multi-physics computational
  model for solving coupled cardiac electromechanics and vascular
  haemodynamics},}\ }\href@noop {} {\bibfield  {journal} {\bibinfo  {journal}
  {arXiv preprint arXiv:2411.11797}\ } (\bibinfo {year} {2024})}\BibitemShut
  {NoStop}%
\bibitem [{\citenamefont {Li}\ \emph {et~al.}(2021{\natexlab{b}})\citenamefont
  {Li}, \citenamefont {Liu-Schiaffini}, \citenamefont {Kovachki}, \citenamefont
  {Liu}, \citenamefont {Azizzadenesheli}, \citenamefont {Bhattacharya},
  \citenamefont {Stuart},\ and\ \citenamefont {Anandkumar}}]{li2021learning}%
  \BibitemOpen
  \bibfield  {author} {\bibinfo {author} {\bibfnamefont {Z.}~\bibnamefont
  {Li}}, \bibinfo {author} {\bibfnamefont {M.}~\bibnamefont {Liu-Schiaffini}},
  \bibinfo {author} {\bibfnamefont {N.}~\bibnamefont {Kovachki}}, \bibinfo
  {author} {\bibfnamefont {B.}~\bibnamefont {Liu}}, \bibinfo {author}
  {\bibfnamefont {K.}~\bibnamefont {Azizzadenesheli}}, \bibinfo {author}
  {\bibfnamefont {K.}~\bibnamefont {Bhattacharya}}, \bibinfo {author}
  {\bibfnamefont {A.}~\bibnamefont {Stuart}},\ and\ \bibinfo {author}
  {\bibfnamefont {A.}~\bibnamefont {Anandkumar}},\ }\bibfield  {title}
  {\enquote {\bibinfo {title} {Learning dissipative dynamics in chaotic
  systems},}\ }\href@noop {} {\bibfield  {journal} {\bibinfo  {journal} {arXiv
  preprint arXiv:2106.06898}\ } (\bibinfo {year}
  {2021}{\natexlab{b}})}\BibitemShut {NoStop}%
\bibitem [{\citenamefont {Shu}, \citenamefont {Li},\ and\ \citenamefont
  {Farimani}(2023)}]{shu2023physics}%
  \BibitemOpen
  \bibfield  {author} {\bibinfo {author} {\bibfnamefont {D.}~\bibnamefont
  {Shu}}, \bibinfo {author} {\bibfnamefont {Z.}~\bibnamefont {Li}},\ and\
  \bibinfo {author} {\bibfnamefont {A.~B.}\ \bibnamefont {Farimani}},\
  }\bibfield  {title} {\enquote {\bibinfo {title} {A physics-informed diffusion
  model for high-fidelity flow field reconstruction},}\ }\href@noop {}
  {\bibfield  {journal} {\bibinfo  {journal} {Journal of Computational
  Physics}\ }\textbf {\bibinfo {volume} {478}},\ \bibinfo {pages} {111972}
  (\bibinfo {year} {2023})}\BibitemShut {NoStop}%
\bibitem [{\citenamefont {Chihaoui}, \citenamefont {Lemkhenter},\ and\
  \citenamefont {Favaro}(2024)}]{chihaoui2024bird}%
  \BibitemOpen
  \bibfield  {author} {\bibinfo {author} {\bibfnamefont {H.}~\bibnamefont
  {Chihaoui}}, \bibinfo {author} {\bibfnamefont {A.}~\bibnamefont
  {Lemkhenter}},\ and\ \bibinfo {author} {\bibfnamefont {P.}~\bibnamefont
  {Favaro}},\ }\bibfield  {title} {\enquote {\bibinfo {title} {Blind image
  restoration via fast diffusion inversion},}\ }\href@noop {} {\bibfield
  {journal} {\bibinfo  {journal} {Advances in Neural Information Processing
  Systems}\ }\textbf {\bibinfo {volume} {37}},\ \bibinfo {pages} {34513--34532}
  (\bibinfo {year} {2024})}\BibitemShut {NoStop}%
\bibitem [{\citenamefont {Wang}\ \emph {et~al.}(2024)\citenamefont {Wang},
  \citenamefont {Yang}, \citenamefont {Chen}, \citenamefont {Wang},
  \citenamefont {Guo}, \citenamefont {Chau}, \citenamefont {Liu}, \citenamefont
  {Qiao}, \citenamefont {Kot},\ and\ \citenamefont {Wen}}]{wang2023sinsr}%
  \BibitemOpen
  \bibfield  {author} {\bibinfo {author} {\bibfnamefont {Y.}~\bibnamefont
  {Wang}}, \bibinfo {author} {\bibfnamefont {W.}~\bibnamefont {Yang}}, \bibinfo
  {author} {\bibfnamefont {X.}~\bibnamefont {Chen}}, \bibinfo {author}
  {\bibfnamefont {Y.}~\bibnamefont {Wang}}, \bibinfo {author} {\bibfnamefont
  {L.}~\bibnamefont {Guo}}, \bibinfo {author} {\bibfnamefont {L.-P.}\
  \bibnamefont {Chau}}, \bibinfo {author} {\bibfnamefont {Z.}~\bibnamefont
  {Liu}}, \bibinfo {author} {\bibfnamefont {Y.}~\bibnamefont {Qiao}}, \bibinfo
  {author} {\bibfnamefont {A.~C.}\ \bibnamefont {Kot}},\ and\ \bibinfo {author}
  {\bibfnamefont {B.}~\bibnamefont {Wen}},\ }\bibfield  {title} {\enquote
  {\bibinfo {title} {Sinsr: diffusion-based image super-resolution in a single
  step},}\ }in\ \href@noop {} {\emph {\bibinfo {booktitle} {Proceedings of the
  IEEE/CVF conference on computer vision and pattern recognition}}}\ (\bibinfo
  {year} {2024})\ pp.\ \bibinfo {pages} {25796--25805}\BibitemShut {NoStop}%
\bibitem [{\citenamefont {Song}, \citenamefont {Meng},\ and\ \citenamefont
  {Ermon}(2020)}]{song2022ddim}%
  \BibitemOpen
  \bibfield  {author} {\bibinfo {author} {\bibfnamefont {J.}~\bibnamefont
  {Song}}, \bibinfo {author} {\bibfnamefont {C.}~\bibnamefont {Meng}},\ and\
  \bibinfo {author} {\bibfnamefont {S.}~\bibnamefont {Ermon}},\ }\bibfield
  {title} {\enquote {\bibinfo {title} {Denoising diffusion implicit models},}\
  }\href@noop {} {\bibfield  {journal} {\bibinfo  {journal} {arXiv preprint
  arXiv:2010.02502}\ } (\bibinfo {year} {2020})}\BibitemShut {NoStop}%
\bibitem [{\citenamefont {Succi}(2001)}]{succi2001lattice}%
  \BibitemOpen
  \bibfield  {author} {\bibinfo {author} {\bibfnamefont {S.}~\bibnamefont
  {Succi}},\ }\href@noop {} {\emph {\bibinfo {title} {The lattice {B}oltzmann
  equation: for fluid dynamics and beyond}}}\ (\bibinfo  {publisher} {{Oxford
  University Press}},\ \bibinfo {year} {2001})\BibitemShut {NoStop}%
\bibitem [{\citenamefont {Smagorinsky}(1963)}]{smagorinsky1963general}%
  \BibitemOpen
  \bibfield  {author} {\bibinfo {author} {\bibfnamefont {J.}~\bibnamefont
  {Smagorinsky}},\ }\bibfield  {title} {\enquote {\bibinfo {title} {General
  circulation experiments with the primitive equations: I. the basic
  experiment},}\ }\href@noop {} {\bibfield  {journal} {\bibinfo  {journal}
  {Monthly weather review}\ }\textbf {\bibinfo {volume} {91}},\ \bibinfo
  {pages} {99--164} (\bibinfo {year} {1963})}\BibitemShut {NoStop}%
\bibitem [{\citenamefont {Koda}\ and\ \citenamefont
  {Lien}(2015)}]{koda2015lattice}%
  \BibitemOpen
  \bibfield  {author} {\bibinfo {author} {\bibfnamefont {Y.}~\bibnamefont
  {Koda}}\ and\ \bibinfo {author} {\bibfnamefont {F.-S.}\ \bibnamefont
  {Lien}},\ }\bibfield  {title} {\enquote {\bibinfo {title} {The lattice
  {B}oltzmann method implemented on the gpu to simulate the turbulent flow over
  a square cylinder confined in a channel},}\ }\href@noop {} {\bibfield
  {journal} {\bibinfo  {journal} {Flow, Turbulence and Combustion}\ }\textbf
  {\bibinfo {volume} {94}},\ \bibinfo {pages} {495--512} (\bibinfo {year}
  {2015})}\BibitemShut {NoStop}%
\end{thebibliography}%

\section{Contributions}
X.Xue, S.Wang conceived the initial idea for this research. X.Xue designed and conducted the numerical simulations. X.Xue, T.Yang M.F.P. ten Eikelder accomplished the theoretical framework. X.Xue, T.Yang, M.Gao, S.Wang, L.Pan, M.Wang, K.Zhu,  performed machine learning training and ablation studies. M.Wang, X.Xue performed Quantum machine learning low-resolution auto-regressive model. X.Xue, T.Yang, M.Gao, J.Li performed the results analysis. P.V.Coveney provided supervision. P.V.Coveney and X.Xue acquired funding and provided access to computing resources. All authors contributed to writing the paper. 

\section{Acknowledgements}
P.V.C. and X.X. acknowledge funding support from the European Commission CompBioMed Centre of Excellence (Grant No. 675451 and 823712). Support from the UK Engineering and Physical Sciences Research Council under the following projects ``UK Consortium on Mesoscale Engineering Sciences (UKCOMES)'' (Grant No.EP/R029598/1) and ``Software Environment for Actionable and VVUQ-evaluated Exascale Applications (SEAVEA)'' (Grant No. EP/W007711/1) is gratefully acknowledged. P.V.C. and X.X. acknowledge the 2024-2025 DOE INCITE award for computational resources on supercomputers at the Oak Ridge Leadership Computing Facility under the ``COMPBIO3'' project. J. Li acknowledges Australian Research Council Project DE26010206. P.V.C. and X.X. acknowledge the use of resources provided by the Isambard-AI National AI Research Resource (AIRR). Isambard-AI is operated by the University of Bristol and is funded by the UK Government’s Department for Science, Innovation and Technology (DSIT) via UK Research and Innovation, and the Science and Technology Facilities Council [ST/AIRR/I-A-I/1023]. This work was supported under project proposals ``GenFLOW" and ``GenFLOW2''; and the Science and Technology Facilities Council [ST/AIRR/I-A-I/1023]. P.V.C and X.X. thank the Leibniz Supercomputing Centre (LRZ) for access to the BEAST GPU cluster. P.V.C, M.D.W. and X.X. also gratefully acknowledge IQM Quantum Computers for providing access to superconducting quantum processors. X. X. gratefully acknowledges the helpful discussions with Tianyi Li during his visit to Rome.

% \section{Competing of interests}
% The authors declare no competing interests.
% %\section{Material availability}
%The training code and data set for sparse data and dense data can be found at the University College London Centre for Computational Science GitHub page: \url{https://github.com/UCL-CCS/PINN-WM-LBM}.
\clearpage
%%%%%%%%% FIG 1 %%%%%%%%%%%%%%

\end{document}